\documentclass[nojss]{jss}

\usepackage{thumbpdf,lmodern,framed,amsmath}

\newcommand{\class}[1]{`\code{#1}'}
\newcommand{\fct}[1]{\code{\detokenize{#1}()}}

\author{Samuel L. Brilleman\\Monash University, Melbourne, Australia
   \And Eren M. Elci\\Bayer AG, Berlin, Germany
   \AND Jacqueline Buros Novik\\Generable Inc., New York, USA
   \And Rory Wolfe\\Monash University, Melbourne, Australia}
\Plainauthor{Samuel L. Brilleman,
             Eren M. Elci,
             Jacqueline Buros Novik,
             Rory Wolfe}
             
\title{Bayesian Survival Analysis Using the \pkg{rstanarm} \proglang{R} Package}
\Plaintitle{Bayesian Survival Analysis Using the rstanarm R Package}
\Shorttitle{Bayesian Survival Analysis Using \pkg{rstanarm}}

\Abstract{
Survival data is encountered in a range of disciplines, most notably health and medical research. Although Bayesian approaches to the analysis of survival data can provide a number of benefits, they are less widely used than classical (e.g. likelihood-based) approaches. This may be in part due to a relative absence of user-friendly implementations of Bayesian survival models. In this article we describe how the \pkg{rstanarm} \proglang{R} package can be used to fit a wide range of Bayesian survival models. The \pkg{rstanarm} package facilitates Bayesian regression modelling by providing a user-friendly interface (users specify their model using customary \proglang{R} formula syntax and data frames) and using the \proglang{Stan} software (a \proglang{C++} library for Bayesian inference) for the back-end estimation. The suite of models that can be estimated using \pkg{rstanarm} is broad and includes generalised linear models (GLMs), generalised linear mixed models (GLMMs), generalised additive models (GAMs) and more. In this article we focus only on the survival modelling functionality. This includes standard parametric (exponential, Weibull, Gompertz) and flexible parametric (spline-based) hazard models, as well as standard parametric accelerated failure time (AFT) models. All types of censoring (left, right, interval) are allowed, as is delayed entry (left truncation), time-varying covariates, time-varying effects, and frailty effects. We demonstrate the functionality through worked examples. We anticipate these implementations will increase the uptake of Bayesian survival analysis in applied research.\\
\medskip
\textbf{Note:} At the time of publishing this working paper the survival analysis functionality in \pkg{rstanarm} was only available on the development branch found at \url{https://github.com/stan-dev/rstanarm/tree/feature/survival}. Hopefully by the time you are reading this, the functionality will be available in the stable release on the Comprehensive R Archive Network (CRAN) at \url{https://CRAN.R-project.org/package=rstanarm}.
}

\Keywords{survival, time-to-event, Bayesian, \proglang{R}, \proglang{Stan}, \pkg{rstanarm}}
\Plainkeywords{survival, time-to-event, Bayesian, R, Stan, rstanarm}

\Address{
  Samuel L. Brilleman\\
  School of Public Health and Preventive Medicine\\
  Monash University\\
  553 St Kilda Road, Melbourne\\
  Victoria 3004\\
  Australia\\
  E-mail: \email{sam.brilleman@monash.edu}\\
  URL: \url{http://www.sambrilleman.com/}
}

\begin{document}



\section{Introduction} \label{sec:intro}

Survival (or time-to-event) analysis is concerned with the analysis of an outcome variable that corresponds to the time from some defined baseline until an event of interest occurs. The methodology is used in a range of disciplines where it is known by a variety of different names. These include survival analysis (medicine), duration analysis (economics), reliability analysis (engineering), and event history analysis (sociology). Survival analyses are particularly common in health and medical research, where a classic example of survival outcome data is the time from diagnosis of a disease until the occurrence of death.

In standard survival analysis, one event time is measured for each observational unit. In practice however that event time may be unobserved due to left, right, or interval censoring, in which case the event time is only known to have occurred within the relevant censoring interval. The combined aspects of time and censoring make survival analysis methodology distinct from many other regression modelling approaches. 

There are two common approaches to modelling survival data. The first is to model the instantaneous rate of the event (known as the hazard) as a function of time. This includes the class of models known as proportional and non-proportional hazards regression models. The second is to model the event time itself. This includes the class of models known as accelerated failure time (AFT) models. Under both of these modelling frameworks a number of extensions have been proposed. For instance the handling of recurrent events, competing events, clustered survival data, cure models, and more. More recently, methods for modelling both longitudinal (e.g. a repeatedly measured biomarker) and survival data have become increasingly popular.

To date, much of the software developed for survival analysis has been based on maximum likelihood or partial likelihood estimation methods. This is in part due to the popularity of the Cox model which is based on a partial likelihood approach that does not require any significant computing resources. Bayesian approaches on the other hand have received much less attention. However, the benefits of Bayesian inference (e.g. the opportunity to make probability statements about parameters, more natural handling of group-specific parameters, better small sample properties, and the ease with which uncertainty can be quantified in predicted quantities) apply just as readily to survival analysis as they do to other areas of statistical modelling and inference \citep{Dunson:2001}.

To our knowledge there are few general purpose Bayesian survival analysis packages for the \proglang{R} software. Perhaps the most extensive package currently available is the \pkg{spBayesSurv} \proglang{R} package \citep{Zhou:2018}. It focuses on Bayesian spatial survival modelling but can also be used for non-spatial survival data. It accommodates all forms of censoring and models can be formulated on a proportional hazards, proportional odds, or AFT scale. Spatial information is handled through random effects. Similarly, the \pkg{spatsurv} \proglang{R} package \citep{Taylor:2017} allows for Bayesian modelling of spatial or non-spatial survival data, with spatial dependencies handled through random effects. However the \pkg{spatsurv} package is limited to just proportional hazards. Both \pkg{spBayesSurv} and \pkg{spatsurv} allow flexible parametric (e.g. smooth) or non-parametric modelling of the baseline hazard or baseline survival function. However one limitation is that neither currently allows for time-varying effects of covariates (e.g. non-proportional hazards).

\proglang{Stan} is a \proglang{C++} library that provides a powerful platform for statistical modelling \citep{Carpenter:2017}. \proglang{Stan} has its own programming language for defining statistical models and interfaces with a number of mainstream statistical software packages to facilitate pre-processing of data and post-estimation inference. Two of the most popular \proglang{Stan} interfaces are available in \proglang{R} (\pkg{RStan}) and \proglang{Python} (\pkg{PyStan}), however others exist for \proglang{Julia} (\pkg{Stan.jl}), \proglang{MATLAB} (\pkg{MatlabStan}), \proglang{Stata} (\pkg{StataStan}), \proglang{Scala} (\pkg{ScalaStan}), \proglang{Mathematica} (\pkg{MathematicaStan}), and the command line (\pkg{CmdStan}).

Regardless of the chosen interface, \proglang{Stan} allows for estimation of statistical models using optimisation, approximate Bayesian inference, or full Bayesian inference. Full Bayesian inference in \proglang{Stan} is based on a specific implementation of Hamiltonian Monte Carlo known as the No-U-Turn Sampler (NUTS) \citep{Hoffman:2014}.

Although \proglang{Stan} is an extremely flexible and powerful software for statistical modelling, it has a relatively high barrier to entry when considered by applied researchers. This is because it requires the user to learn the \proglang{Stan} programming language in order to define their statistical model. For this reason, several high-level interfaces have been developed. These high-level interfaces shield the user from any \proglang{Stan} code as well as provide useful tools to facilitate model checking, model inference, and generating predictions.

One of the most popular high-level interfaces for \proglang{Stan} is the \pkg{rstanarm} \proglang{R} package \citep{Goodrich:2018}, available from the Comprehensive R Archive Network (CRAN) at \url{https://CRAN.R-project.org/package=rstanarm}. The \pkg{rstanarm} \proglang{R} package allows users to fit a broad range of regression models using customary \proglang{R} formula syntax and data frames. The user is not required to write any \proglang{Stan} code themselves, yet \code{Stan} is used for the back-end estimation. The \pkg{rstanarm} package includes functionality for fitting generalised linear models (GLMs), generalised linear mixed models (GLMMs), generalised additive models (GAMs), survival models, and more. Describing all of the regression modelling functionality in \pkg{rstanarm} is well beyond the scope of a single article and there is a series of vignettes that attempt that task. Instead, in this article we focus only on describing and demonstrating the survival modelling functionality in \pkg{rstanarm}.

Our article is therefore structured as follows. In Sections \ref{sec:modelling}, \ref{sec:estimation}, and \ref{sec:prediction} we describe the modelling, estimation, and prediction frameworks underpinning survival models in \pkg{rstanarm}. In Section \ref{sec:implementation} we describe implementation of the methods as they exist within the package. In Section \ref{sec:usage} we demonstrate usage of the package through a series of examples. In Section \ref{sec:summary} we close with a discussion.



\section{Modelling framework} \label{sec:modelling}

\subsection{Data and notation}

We assume that a true event time for individual $i$ ($i = 1,...,N$) exists and can be denoted $T_i^*$. However, in practice $T_i^*$ may not be observed due to left, right, or interval censoring. We therefore observe outcome data $\mathcal{D}_i = \{T_i, T_i^U, T_i^E, d_i\}$ for individual $i$ where:

\begin{itemize}

\item $T_i$ denotes the observed event or censoring time;

\item $T_i^U$ denotes the observed upper limit for interval censored individuals;

\item $T_i^E$ denotes the observed entry time (the time at which an individual became at risk of experiencing the event); and

\item $d_i \in \{0,1,2,3\}$ denotes an event indicator taking value 0 if individual $i$ was right censored (i.e. $T_i^* > T_i$), value 1 if individual $i$ was uncensored (i.e. $T_i^* = T_i$), value 2 if individual $i$ was left censored (i.e. $T_i^* < T_i$), or value 3 if individual $i$ was interval censored (i.e. $T_i < T_i^* < T_i^U$).

\end{itemize}

\subsubsection{Hazard, cumulative hazard, and survival}

There are three key quantities of interest in standard survival analysis: the hazard rate, the cumulative hazard, and the survival probability. It is these quantities that are used to form the likelihood function for the survival models described in later sections.

The hazard is the instantaneous rate of occurrence for the event at time $t$. Mathematically, it is defined as:
\begin{equation} \label{eq:hazdef}
\begin{split}
h_i(t) = \lim_{\Delta t \to 0} 
  \frac{P(t \leq T_i^* < t + \Delta t | T_i^* > t)}{\Delta t}
\end{split}
\end{equation}
where $\Delta t$ is the width of some small time interval.

The numerator in Equation (\ref{eq:hazdef}) is the conditional probability of the individual experiencing the event during the time interval $[t, t + \Delta t)$, given that they were still at risk of the event at time $t$. The denominator in Equation (\ref{eq:hazdef}) converts the conditional probability to a rate per unit of time. As $\Delta t$ approaches the limit, the width of the interval approaches zero and the instantaneous event rate is obtained.

The cumulative hazard is defined as:
\begin{equation} \label{eq:chazdef}
\begin{split}
H_i(t) = \int_{u=0}^t h_i(u) du
\end{split}
\end{equation}
and the survival probability is defined as:
\begin{equation} \label{eq:survdef}
\begin{split}
S_i(t) = \exp \left( -H_i(t) \right) = \exp \left( -\int_{u=0}^t h_i(u) du \right)
\end{split}
\end{equation}

It can be seen here that in the standard survival analysis setting -- where there is one event type of interest (i.e. no competing events) -- there is a one-to-one relationship between each of the hazard, the cumulative hazard, and the survival probability.

\subsubsection{Delayed entry}

Delayed entry (also known as left truncation) occurs when an individual is not at risk of the event until some time $t > 0$. As previously described we use $T_i^E$ to denote the entry time at which the individual becomes at risk. A common situation where delayed entry occurs is when age is used as the time scale. With age as the time scale it is likely that our study will only be concerned with the observation of individuals starting from some time (i.e. age) $t > 0$.

To allow for delayed entry we essentially want to work with a conditional survival probability:
\begin{equation} \label{eq:survdef}
\begin{split}
S_i \left(t \mid T_i^E > 0 \right) = \frac{S_i(t)}{S_i \left( T_i^E \right)}
\end{split}
\end{equation}

Here the survival probability is evaluated conditional on the individual having survived up to the entry time. We will see this approach used in Section \ref{sec:loglikelihood} where we define the log likelihood for our survival model.

\subsection{Model formulations} \label{sec:modelformulations}

Our modelling approaches are twofold. First, we define a class of models on the hazard scale. This includes both proportional and non-proportional hazard regression models. Second, we define a class of models on the scale of the survival time. These are often known as accelerated failure time (AFT) models and can include both time-fixed and time-varying acceleration factors.

These two classes of models and their respective features are described in the following sections.

\subsubsection{Hazard scale models}

Under a hazard scale formulation, we model the hazard of the event for individual $i$ at time $t$ using the regression model:
\begin{equation}
\begin{split}
h_i(t) = h_0(t) \exp \left( \eta_i(t) \right)
\end{split}
\end{equation}
where $h_0(t)$ is the baseline hazard (i.e. the hazard for an individual with all covariates set equal to zero) at time $t$, and $\eta_i(t)$ denotes the linear predictor evaluated for individual $i$ at time $t$.

For full generality we allow the linear predictor to be time-varying. That is, it may be a function of time-varying covariates and/or time-varying coefficients (e.g. a time-varying hazard ratio). However, if there are no time-varying covariates or time-varying coefficients in the model, then the linear predictor reduces to a time-fixed quantity and the definition of the hazard function reduces to:
\begin{equation}
\begin{split}
h_i(t) = h_0(t) \exp \left( \eta_i \right)
\end{split}
\end{equation}
where the linear predictor $\eta_i$ is no longer a function of time. We describe the linear predictor in detail in later sections.

Different distributional assumptions can be made for the baseline hazard $h_0(t)$ and affect how the baseline hazard changes as a function of time. The \pkg{rstanarm} package currently accommodates several standard parametric distributions for the baseline hazard (exponential, Weibull, Gompertz) as well as more flexible approaches that directly model the baseline hazard as a piecewise or smooth function of time using splines.

The following describes the baseline hazards that are currently implemented in the \pkg{rstanarm} package.

\paragraph{M-splines model (the default):}

Let $M_{l}(t; \boldsymbol{k}, \delta)$ denote the $l^{\text{th}}$ $(l = 1,...,L)$ basis term for a degree $\delta$ M-spline function evaluated at a vector of knot locations $\boldsymbol{k} = \{k_{1},...,k_{J}\}$, and $\gamma_{l}$ denote the $l^{\text{th}}$ M-spline coefficient. We then have:
\begin{equation}
h_i(t) = \sum_{l=1}^{L} \gamma_{l} M_{l}(t; \boldsymbol{k}, \delta) \exp ( \eta_i(t) )
\end{equation}

The M-spline basis is evaluated using the method described in \cite{Ramsay:1988} and implemented in the \pkg{splines2} \proglang{R} package \citep{Wang:2018}.

To ensure that the hazard function $h_i(t)$ is not constrained to zero at the origin (i.e. when $t$ approaches 0) the M-spline basis incorporates an intercept. To ensure identifiability of both the M-spline coefficients and the intercept in the linear predictor we constrain the M-spline coefficients to a simplex, that is, $\sum_{l=1}^L{\gamma_l} = 1$.

The default degree in \pkg{rstanarm} is $\delta = 3$ (i.e. cubic M-splines) such that the baseline hazard can be modelled as a flexible and smooth function of time, however this can be changed by the user. It is worthwhile noting that setting $\delta = 0$ is treated as a special case that corresponds to a piecewise constant baseline hazard.

\paragraph{Exponential model:}

For scale parameter $\lambda_i(t) = \exp ( \eta_i(t) )$ we have:
\begin{equation}
h_i(t) = \lambda_i(t)
\end{equation}

In the case where the linear predictor is not time-varying, the exponential model leads to a hazard rate that is constant over time.

\paragraph{Weibull model:}

For scale parameter $\lambda_i(t) = \exp ( \eta_i(t) )$ and shape parameter $\gamma > 0$ we have:
\begin{equation}
h_i(t) = \gamma t^{\gamma-1} \lambda_i(t)
\end{equation}

In the case where the linear predictor is not time-varying, the Weibull model leads to a hazard rate that is monotonically increasing or monotonically decreasing over time. In the special case where $\gamma = 1$ it reduces to the exponential model.

\paragraph{Gompertz model:}

For shape parameter $\lambda_i(t) = \exp ( \eta_i(t) )$ and scale parameter $\gamma > 0$ we have:
\begin{equation}
h_i(t) = \exp(\gamma t) \lambda_i(t)
\end{equation}

\paragraph{B-splines model (for the log baseline hazard):}

Let $B_{l}(t; \boldsymbol{k}, \delta)$ denote the $l^{\text{th}}$ $(l = 1,...,L)$ basis term for a degree $\delta$ B-spline function evaluated at a vector of knot locations $\boldsymbol{k} = \{k_{1},...,k_{J}\}$, and $\gamma_{l}$ denote the $l^{\text{th}}$ B-spline coefficient. We then have:
\begin{equation}
h_i(t) = \exp \left( \sum_{l=1}^{L} \gamma_{l} B_{l}(t; \boldsymbol{k}, \delta) + \eta_i(t) \right)
\end{equation}
The B-spline basis is calculated using the method implemented in the \pkg{splines2} \proglang{R} package \citep{Wang:2018}. The B-spline basis does not require an intercept and therefore does not include one; any constant shift in the log hazard is fully captured via an intercept in the linear predictor. By default cubic B-splines are used (i.e. $\delta = 3$) and these allow the log baseline hazard to be modelled as a smooth function of time.

\subsubsection{Accelerated failure time (AFT) models}

Under an AFT formulation we model the survival probability for individual $i$ at time $t$ using the regression model \citep{Hougaard:1999}:
\begin{equation} \label{eq:aftform-surv}
\begin{split}
S_i(t) = S_0 \left( \int_{u=0}^t \exp \left( - \eta_i(u) \right) du \right)
\end{split}
\end{equation}
where $S_0(t)$ is the baseline survival probability at time $t$, and $\eta_i(t)$ denotes the linear predictor evaluated for individual $i$ at time $t$. For full generality we again allow the linear predictor to be time-varying. This also leads to a corresponding general expression for the hazard function \citep{Hougaard:1999} as follows:
\begin{align} \label{eq:aftform-haz}
\begin{split}
h_i(t) = \exp \left(-\eta_i(t) \right) h_0 \left( \int_{u=0}^t \exp \left( - \eta_i(u) \right) du \right)
\end{split}
\end{align}

If there are no time-varying covariates or time-varying coefficients in the model, then the definition of the survival probability reduces to:
\begin{equation}
\begin{split}
S_i(t) = S_0 \left( t \exp \left( - \eta_i \right) \right)
\end{split}
\end{equation}
and for the hazard:
\begin{align}
\begin{split}
h_i(t) = \exp \left( -\eta_i \right) h_0 \left( t \exp \left( - \eta_i \right) \right)
\end{split}
\end{align}

Different distributional assumptions can be made for how the baseline survival probability $S_0(t)$ changes as a function of time. The \pkg{rstanarm} package currently accommodates two standard parametric distributions (exponential, Weibull) although others may be added in the future. The current distributions are implemented as follows.

\paragraph{Exponential model:}

When the linear predictor is time-varying we have:
\begin{equation}
S_i(t) = \exp \left( - \int_{u=0}^t \exp ( -\eta_i(u) ) du \right)
\end{equation}
and when the linear predictor is time-fixed we have:
\begin{equation}
S_i(t) = \exp \left( - t \lambda_i \right)
\end{equation}
for scale parameter $\lambda_i = \exp ( -\eta_i )$.

\paragraph{Weibull model:}

When the linear predictor is time-varying we have:
\begin{equation}
S_i(t) = \exp \left( - \left( \int_{u=0}^t \exp ( -\eta_i(u) ) du \right)^{\gamma} \right)
\end{equation}
for shape parameter $\gamma > 0$ and when the linear predictor is time-fixed we have:
\begin{equation}
S_i(t) = \exp \left( - t^{\gamma} \lambda_i \right)
\end{equation}
for scale parameter $\lambda_i = \exp ( -\gamma \eta_i )$ and shape parameter $\gamma > 0$.

\subsection{Linear predictor} \label{sec:linearpredictor}

Under all of the previous model formulations our linear predictor can be defined as:
\begin{equation} \label{eq:eta}
\begin{split}
\eta_i(t) = \boldsymbol{\beta}^T(t) \boldsymbol{X}_i(t)
\end{split}
\end{equation}
where $\boldsymbol{X}_i(t) = [1, x_{i1}(t), ..., x_{iP}(t) ]$ denotes a vector of covariates with $x_{ip}(t)$ denoting the observed value of $p^{th}$ $(p = 1,...,P)$ covariate for the $i^{th}$ $(i=1,...,N)$ individual at time $t$, and $\boldsymbol{\beta}(t) = [ \beta_0, \beta_1(t), ... , \beta_P(t) ]$ denotes a vector of parameters with $\beta_0$ denoting an intercept parameter and $\beta_p(t)$ denoting the possibly time-varying coefficient for the $p^{th}$ covariate.

\subsubsection{Hazard ratios}

Under a hazard scale formulation the quantity $\exp \left( \beta_p(t) \right)$ is referred to as a \emph{hazard ratio}.

The hazard ratio quantifies the relative increase in the hazard that is associated with a unit-increase in the relevant covariate, $x_{ip}$, assuming that all other covariates in the model are held constant. For instance, a hazard ratio of 2 means that a unit-increase in the covariate leads to a doubling in the hazard (i.e. the instantaneous rate) of the event.

\subsubsection{Acceleration factors and survival time ratios}

Under an AFT formulation the quantity $\exp \left( - \beta_p(t) \right)$ is referred to as an \emph{acceleration factor} and the quantity $\exp \left( \beta_p(t) \right)$ is referred to as a \emph{survival time ratio}.

The acceleration factor quantifies the acceleration (or deceleration) of the event process that is associated with a unit-increase in the relevant covariate, $x_{ip}$. For instance, an acceleration factor of 0.5 means that a unit-increase in the covariate corresponds to approaching the event at half the speed.

The survival time ratio is interpreted as the increase (or decrease) in the expected survival time that is associated with a unit-increase in the relevant covariate, $x_{ip}$. For instance, a survival time ratio of 2 (which is equivalent to an acceleration factor of 0.5) means that a unit-increase in the covariate leads to an doubling in the expected survival time.

Note that the survival time ratio is a simple reparameterisation of the acceleration factor. Specifically, the survival time ratio is equal to the reciprocal of the acceleration factor. The survival time ratio and the acceleration factor therefore provide alternative interpretations for the same effect of the same covariate.

\subsubsection{Time-fixed vs time-varying effects}

Under either a hazard scale or AFT formulation the coefficient $\beta_p(t)$ can be treated as a time-fixed or time-varying quantity.

When $\beta_p(t)$ is treated as a time-fixed quantity we have:
\begin{equation}
\begin{split}
\beta_p(t) = \theta_{p0}
\end{split}
\end{equation}
such that $\theta_{p0}$ is a time-fixed log hazard ratio (or log survival time ratio). On the hazard scale this is equivalent to assuming proportional hazards, whilst on the AFT scale it is equivalent to assuming a time-fixed acceleration factor.

When $\beta_p(t)$ is treated as a time-varying quantity we refer to it as a time-varying effect because the effect of the covariate is allowed to change as a function of time. On the hazard scale this leads to non-proportional hazards, whilst on the AFT scale it leads to time-varying acceleration factors.

When $\beta_p(t)$ is time-varying we must determine how we wish to model it. In \pkg{rstanarm} the default is to use B-splines such that:
\begin{equation}
\begin{split}
\beta_p(t) = \theta_{p0} + \sum_{l=1}^{L} \theta_{pl} B_{l}(t; \boldsymbol{k}, \delta)
\end{split}
\end{equation}
where $\theta_{p0}$ is a constant, $B_{l}(t; \boldsymbol{k}, \delta)$ is the $l^{\text{th}}$ $(l = 1,...,L)$ basis term for a degree $\delta$ B-spline function evaluated at a vector of knot locations $\boldsymbol{k} = \{k_{1},...,k_{J}\}$, and $\theta_{pl}$ is the $l^{\text{th}}$ B-spline coefficient. By default cubic B-splines are used (i.e. $\delta = 3$). These allow the log hazard ratio (or log survival time ratio) to be modelled as a smooth function of time. 

However an alternative is to model $\beta_p(t)$ using a piecewise constant function:
\begin{equation}
\begin{split}
\beta_p(t) = \theta_{p0} + \sum_{l=1}^{L} \theta_{pl} I(k_{l+1} < t \leq k_{l+2})
\end{split}
\end{equation}
where $I(x)$ is an indicator function taking value 1 if $x$ is true and 0 otherwise, $\theta_{p0}$ is a constant corresponding to the log hazard ratio (or log survival time ratio for AFT models) in the first time interval, $\theta_{pl}$ is the deviation in the log hazard ratio (or log survival time ratio) between the first and $(l+1)^\text{th}$ $(l = 1,...,L)$ time interval, and $\boldsymbol{k} = \{k_{1},...,k_{J}\}$ is a sequence of knot locations (i.e. break points) that includes the lower and upper boundary knots. This allows the log hazard ratio (or log survival time ratio) to be modelled as a piecewise constant function of time.

Note that we have dropped the subscript $p$ from the knot locations $\boldsymbol{k}$ and degree $\delta$ discussed above. This is just for simplicity of the notation. In fact, if a model has a time-varying effect estimated for more than one covariate, then each of these can be modelled using different knot locations and/or degree if the user desires. These knot locations and/or degree can also differ from those used for modelling the baseline or log baseline hazard described previously in Section \ref{sec:modelformulations}.

\subsubsection{Relationship between proportional hazards and AFT models} \label{sec:aft-haz-relationship}

As shown in Section \ref{sec:modelformulations} some baseline distributions can be parameterised as either a proportional hazards or an AFT model. In \pkg{rstanarm} this currently includes the exponential and Weibull models. One can therefore transform the estimates from an exponential or Weibull proportional hazards model to get the estimates that would be obtained under an exponential or Weibull AFT parameterisation.

Specifically, the following relationship applies for the exponential model:
\begin{equation}
\begin{split}
\beta_0 & = - \beta_0^* \\
\beta_p & = - \beta_p^*  
\end{split}
\end{equation}
and for the Weibull model:
\begin{equation}
\begin{split}
\beta_0 & = -\gamma \beta_0^* \\
\beta_p & = -\gamma \beta_p^*  
\end{split}
\end{equation}
where the unstarred parameters are from the proportional hazards model and the starred ($*$) parameters are from the AFT model. Note however that these relationships only hold in the absence of time-varying effects. This is demonstrated using a real dataset in the example in Section \ref{sec:aftmodel}.

\subsection{Multilevel survival models}

The definition of the linear predictor in Equation \ref{eq:eta} can be extended to allow for shared frailty or other clustering effects.

Suppose that the individuals in our sample belong to a series of clusters. The clusters may represent for instance hospitals, families, or GP clinics. We denote the $i^{th}$ individual ($i = 1,...,N_j$) as a member of the $j^{th}$ cluster ($j = 1,...,J$). Moreover, to indicate the fact that individual $i$ is now a member of cluster $j$ we index the observed data (i.e. event times, event indicator, and covariates) with a subscript $j$, that is $T_{ij}^*$, $\mathcal{D}_{ij} = \{T_{ij}, T_{ij}^U, T_{ij}^E, d_{ij}\}$ and $X_{ij}(t)$, as well as estimated quantities such as the hazard rate, cumulative hazard, survival probability, and linear predictor, that is $h_{ij}(t)$, $H_{ij}(t)$, $S_{ij}(t)$, and $\eta_{ij}(t)$.

To allow for intra-cluster correlation in the event times we include cluster-specific random effects in the linear predictor as follows:
\begin{equation} \label{eq:multileveleta}
\begin{split}
\eta_{ij}(t) = \boldsymbol{\beta}^T \boldsymbol{X}_{ij}(t) + \boldsymbol{b}_{j}^T \boldsymbol{Z}_{ij}
\end{split}
\end{equation}
where $\boldsymbol{Z}_{ij}$ denotes a vector of covariates for the $i^{th}$ individual in the $j^{th}$ cluster, with an associated vector of cluster-specific parameters $\boldsymbol{b}_{j}$. We assume that the cluster-specific parameters are normally distributed such that $\boldsymbol{b}_{j} \sim N(0, \boldsymbol{\Sigma}_{b})$ for some variance-covariance matrix $\boldsymbol{\Sigma}_{b}$. We assume that $\boldsymbol{\Sigma}_{b}$ is unstructured, that is each variance and covariance term is allowed to be different.

In most cases $\boldsymbol{b}_{j}$ will correspond to just a cluster-specific random intercept (often known as a "shared frailty" term) but more complex random effects structures are possible. 

For simplicitly of notation Equation \ref{eq:multileveleta} also assumes just one clustering factor in the model (indexed by $j = 1,...,J$). However it is possible to extend the model to multiple clustering factors. For example, suppose that the $i^{th}$ individual was clustered within the $j^{th}$ hospital that was clustered within the $k^{th}$ geographical region. Then we would have hospital-specific random effects $\boldsymbol{b}_j \sim N(0, \boldsymbol{\Sigma}_{b})$ and region-specific random effects $\boldsymbol{u}_k \sim N(0, \boldsymbol{\Sigma}_{u})$ and assume $\boldsymbol{b}_j$ and $\boldsymbol{u}_k$ are independent for all $(j,k)$. Multiple clustering factors are accommodated as part of the survival modelling functionality in \pkg{rstanarm}.

\section{Estimation framework} \label{sec:estimation}

\subsection{Log posterior}

The log posterior for the $i^{th}$ individual in the $j^{th}$ cluster can be specified as:
\begin{equation}
\begin{split}
\log p(\boldsymbol{\theta}, \boldsymbol{b}_{j} \mid \mathcal{D}_{ij})
  \propto
    \log p(\mathcal{D}_{ij} \mid \boldsymbol{\theta}, \boldsymbol{b}_{j}) +
    \log p(\boldsymbol{b}_{j} \mid \boldsymbol{\theta}) +
    \log p(\boldsymbol{\theta})  
\end{split}
\end{equation}
where $\log p(\mathcal{D}_{ij} \mid \boldsymbol{\theta}, \boldsymbol{b}_{j})$ is the log likelihood for the outcome data, $\log p(\boldsymbol{b}_{j} \mid \boldsymbol{\theta})$ is the log likelihood for the distribution of any cluster-specific parameters (i.e. random effects) when relevant, and $\log p(\boldsymbol{\theta})$ represents the log likelihood for the joint prior distribution across all remaining unknown parameters.

\subsection{Log likelihood} \label{sec:loglikelihood}

Allowing for the three forms of censoring (left, right, and interval censoring) and potential delayed entry (i.e. left truncation) the log likelihood for the survival model takes the form:
\begin{equation} \label{eq:loglik}
\begin{split}
\log p(\mathcal{D}_{ij} \mid \boldsymbol{\theta}, \boldsymbol{b}_{j})
  & =       {I(d_{ij} = 0)} \times \log \left[ S_{ij}(T_{ij})                    \right] \\
  & \quad + {I(d_{ij} = 1)} \times \log \left[ h_{ij}(T_{ij})                    \right] \\
  & \quad + {I(d_{ij} = 1)} \times \log \left[ S_{ij}(T_{ij})                    \right] \\
  & \quad + {I(d_{ij} = 2)} \times \log \left[ 1 - S_{ij}(T_{ij})                \right] \\
  & \quad + {I(d_{ij} = 3)} \times \log \left[ S_{ij}(T_{ij}) - S_{ij}(T_{ij}^U) \right] \\
  & \quad - \log \left[ S_{ij} ( T_{ij}^E ) \right]
\end{split}
\end{equation}
where $I(x)$ is an indicator function taking value 1 if $x$ is true and 0 otherwise. That is, each individual's contribution to the likelihood depends on the type of censoring for their event time. 

The last term on the right hand side of Equation \ref{eq:loglik} accounts for delayed entry. When an individual is at risk from time zero (i.e. no delayed entry) then $T_{ij}^E = 0$ and $S_{ij}(0) = 1$ meaning that the last term disappears from the likelihood.

\subsubsection{Evaluating integrals in the log likelihood}

When the linear predictor is time-fixed there is a closed form expression for both the hazard rate and survival probability in almost all cases (the single exception is when B-splines are used to model the log baseline hazard). When there is a closed form expression for both the hazard rate and survival probability then there is also a closed form expression for the (log) likelihood function. The details of these expressions are given in Appendix \ref{app:haz-parameterisations} (for hazard models) and Appendix \ref{app:aft-parameterisations} (for AFT models).

However, when the linear predictor is time-varying there isn't a closed form expression for the survival probability. Instead, Gauss-Kronrod quadrature with $Q$ nodes is used to approximate the necessary integrals.

For hazard scale models Gauss-Kronrod quadrature is used to evaluate the cumulative hazard, which in turn is used to evaluate the survival probability. Expanding on Equation \ref{eq:survdef} we have:
\begin{equation}
\begin{split}
\int_{u=0}^{T_{ij}} h_{ij}(u) du
  \approx \frac{T_{ij}}{2} \sum_{q=1}^{Q} w_q h_{ij} \left( \frac{T_{ij}(1 + v_q)}{2} \right)
\end{split}
\end{equation}
where $w_q$ and $v_q$, respectively, are the standardised weights and locations ("abscissa") for quadrature node $q$ $(q = 1,...,Q)$ \citep{Laurie:1997}.

For AFT models Gauss-Kronrod quadrature is used to evaluate the cumulative acceleration factor, which in turn is used to evaluate both the survival probability and the hazard rate. Expanding on Equations \ref{eq:aftform-surv} and \ref{eq:aftform-haz} we have:
\begin{equation}
\begin{split}
\int_{u=0}^{T_{ij}} \exp \left( - \eta_{ij}(u) \right) du
  \approx \frac{T_{ij}}{2} \sum_{q=1}^{Q} w_q \exp \left( - \eta_{ij} \left( \frac{T_{ij}(1 + v_q)}{2} \right) \right)
\end{split}
\end{equation}

When quadrature is necessary, the default in \pkg{rstanarm} is to use $Q = 15$ nodes. But the number of nodes can be changed by the user.

\subsection{Prior distributions}

For each of the parameters a number of prior distributions are available. Default choices exist, but the user can explicitly specify the priors if they wish.

\subsubsection{Intercept}

All models include an intercept parameter in the linear predictor ($\beta_0$) which effectively forms part of the baseline hazard. Choices of prior distribution for $\beta_0$ include the normal, t, or Cauchy distributions. The default is a normal distribution with mean 0 and standard deviation of 20.

However it is worth noting that -- internally (but not in the reported parameter estimates) -- the prior is placed on the intercept after centering the predictors at their sample means and after applying a constant shift of $\log \left( \frac{E}{T} \right)$ where $E$ is the total number of events and $T$ is the total follow up time. For instance, the default prior is not centered on an intercept of zero when all predictors are at their sample means, but rather, it is centered on the log crude event rate when all predictors are at their sample means. This is intended to help with numerical stability and sampling, but does not impact on the reported estimates (i.e. the intercept is back-transformed before being returned to the user).

\subsubsection{Regression coefficients}

Choices of prior distribution for the time-fixed regression coefficients $\theta_{p0}$ ($p = 1,...,P$) include normal, t, and Cauchy distributions as well as several shrinkage prior distributions.

Where relevant, the additional coefficients required for estimating a time-varying effect (i.e. the B-spline coefficients or the interval-specific deviations in the piecewise constant function) are given a random walk prior of the form $\theta_{p,1} \sim N(0,1)$ and $\theta_{p,m} \sim N(\theta_{p,m-1},\tau_p)$ for $m = 2,...,M$, where $M$ is the total number of cubic B-spline basis terms. The prior distribution for the hyperparameter $\tau_p$ can be specified by the user and choices include an exponential, half-normal, half-t, or half-Cauchy distribution. Note that lower values of $\tau_p$ lead to a less flexible (i.e. smoother) function for modelling the time-varying effect.

\subsubsection{Auxiliary parameters}

There are several choices of prior distribution for the so-called "auxiliary" parameters related to the baseline hazard (i.e. scalar $\gamma$ for the Weibull and Gompertz models or vector $\boldsymbol{\gamma}$ for the M-spline and B-spline models). These include:

\begin{itemize}

\item a Dirichlet prior distribution for the baseline hazard M-spline coefficients $\boldsymbol{\gamma}$;

\item a half-normal, half-t, half-Cauchy or exponential prior distribution for the Weibull shape parameter $\gamma$;

\item a half-normal, half-t, half-Cauchy or exponential prior distribution for the Gompertz scale parameter $\gamma$; and

\item a normal, t, or Cauchy prior distribution for the log baseline hazard B-spline coefficients $\boldsymbol{\gamma}$.

\end{itemize}

\subsubsection{Covariance matrices}

When a multilevel survival model is estimated there is an unstructured covariance matrix estimated for the random effects. Of course, in the situation where there is just one random effect in the model \code{formula} (e.g. a random intercept or "shared frailty" term) the covariance matrix will reduce to just a single element; i.e. it will be a scalar equal to the variance of the single random effect in the model.

The prior distribution is based on a decomposition of the covariance matrix. The decomposition takes place as follows. The covariance matrix $\boldsymbol{\Sigma}_b$ is decomposed into a correlation matrix $\boldsymbol{\Omega}$ and vector of variances. The vector of variances is then further decomposed into a simplex $\pi$ (i.e. a probability vector summing to 1) and a scalar equal to the sum of the variances. Lastly, the sum of the variances is set equal to the order of the covariance matrix multiplied by the square of a scale parameter (here we denote that scale parameter $\tau$).

The prior distribution for the correlation matrix $\boldsymbol{\Omega}$ is the LKJ distribution \citep{Lewandowski:2009}. It is parameterised through a regularisation parameter $\zeta > 0$. The default is $\zeta = 1$ such that the LKJ prior distribution is jointly uniform over all possible correlation matrices. When $\zeta > 1$ the mode of the LKJ distribution is the identity matrix and as $\zeta$ increases the distribution becomes more sharply peaked at the mode. When $0 < \zeta < 1$ the prior has a trough at the identity matrix.

The prior distribution for the simplex $\boldsymbol{\pi}$ is a symmetric Dirichlet distribution with a single concentration parameter $\phi > 0$. The default is $\phi = 1$ such that the prior is jointly uniform over all possible simplexes. If $\phi > 1$ then the prior mode corresponds to all entries of the simplex being equal (i.e. equal variances for the random effects) and the larger the value of $\phi$ then the more pronounced the mode of the prior. If $0 < \phi < 1$ then the variances are polarised. 

The prior distribution for the scale parameter $\tau$ is a Gamma distribution. The shape and scale parameter for the Gamma distribution are both set equal to 1 by default, however the user can change the value of the shape parameter. The behaviour is such that increasing the shape parameter will help enforce that the trace of $\boldsymbol{\Sigma}_b$ (i.e. sum of the variances of the random effects) be non-zero.

Further details on this implied prior for covariance matrices can be found in the \pkg{rstanarm} documentation and vignettes.

\subsection{Estimation}

Estimation in \pkg{rstanarm} is based on either full Bayesian inference (Hamiltonian Monte Carlo) or approximate Bayesian inference (either mean-field or full-rank variational inference). The default is full Bayesian inference, but the user can change this if they wish. The approximate Bayesian inference algorithms are much faster, but they only provide approximations for the joint posterior distribution and are therefore not recommended for final inference.

Hamiltonian Monte Carlo is a form of Markov chain Monte Carlo (MCMC) in which information about the gradient of the log posterior is used to more efficiently sample from the posterior space. \proglang{Stan} uses a specific implementation of Hamiltonian Monte Carlo known as the No-U-Turn Sampler (NUTS) \citep{Hoffman:2014}. A benefit of NUTS is that the tuning parameters are handled automatically during a "warm-up" phase of the estimation. However the \pkg{rstanarm} modelling functions provide arguments that allow the user to retain control over aspects such as the number of MCMC chains, number of warm-up and sampling iterations, and number of computing cores used.

\section{Prediction framework} \label{sec:prediction}

\subsection{Survival predictions without clustering}

If our survival model does not contain any clustering effects (i.e. it is not a multilevel survival model) then our prediction framework is more straightforward. Let $\mathcal{D} = \{ \mathcal{D}_{i}; i = 1,...,N \}$ denote the entire collection of outcome data in our sample and let $T_{\max} = \max \{ T_{i}, T_{i}^U, T_{i}^E; i = 1,...,N \}$ denote the maximum event or censoring time across all individuals in our sample.

Suppose that for some individual $i^*$ (who may or may not have been in our sample) we have covariate vector $\boldsymbol{x}_{i^*}$. Note that the covariate data must be time-fixed. The predicted probability of being event-free at time $0 < t \leq T_{\max}$, denoted $\hat{S}_{i^*}(t)$, can be generated from the posterior predictive distribution:
\begin{equation}
\begin{split}
p \Big( \hat{S}_{i^*}(t) \mid \boldsymbol{x}_{i^*}, \mathcal{D} \Big) = 
  \int
    p \Big( \hat{S}_{i^*}(t) \mid \boldsymbol{x}_{i^*}, \boldsymbol{\theta} \Big)
    p \Big( \boldsymbol{\theta} \mid \mathcal{D} \Big)
  d \boldsymbol{\theta}
\end{split}
\end{equation}

We approximate this posterior predictive distribution by drawing from $p(\hat{S}_{i^*}(t) \mid \boldsymbol{x}_{i^*}, \boldsymbol{\theta}^{(l)})$ where $\boldsymbol{\theta}^{(l)}$ is the $l^{th}$ $(l = 1,...,L)$ MCMC draw from the posterior distribution $p(\boldsymbol{\theta} \mid \mathcal{D})$. 

\subsection{Survival predictions with clustering}

When there are clustering effects in the model (i.e. multilevel survival models) then our prediction framework requires conditioning on the cluster-specific parameters. Let $\mathcal{D} = \{ \mathcal{D}_{ij}; i = 1,...,N_j, j = 1,...,J \}$ denote the entire collection of outcome data in our sample and let $T_{\max} = \max \{ T_{ij}, T_{ij}^U, T_{ij}^E; i = 1,...,N_j, j = 1,...,J \}$ denote the maximum event or censoring time across all individuals in our sample.

Suppose that for some individual $i^*$ (who may or may not have been in our sample) and who is known to come from cluster $j^*$ (which may or may not have been in our sample) we have covariate vectors $\boldsymbol{x}_{i^*j^*}$ and $\boldsymbol{z}_{i^*j^*}$. Note again that the covariate data is assumed to be time-fixed.

If individual $i^*$ does in fact come from a cluster $j^* = j$ (for some $j \in \{1,...,J\}$) in our sample then the predicted probability of being event-free at time $0 < t \leq T_{\max}$, denoted $S_{i^*j}(t)$, can be generated from the posterior predictive distribution:
\begin{equation}
\begin{split}
p \Big( \hat{S}_{i^*j}(t) \mid \boldsymbol{x}_{i^*j}, \boldsymbol{z}_{i^*j}, \mathcal{D} \Big) = 
  \int
    \int
      p \Big( \hat{S}_{i^*j}(t) \mid \boldsymbol{x}_{i^*j}, \boldsymbol{z}_{i^*j}, \boldsymbol{\theta}, \boldsymbol{b}_j \Big)
      p \Big( \boldsymbol{\theta}, \boldsymbol{b}_j \mid \mathcal{D} \Big)
    d \boldsymbol{b}_j \space d \boldsymbol{\theta}
\end{split}
\end{equation}

Since cluster $j$ was included in our sample data it is easy for us to approximate this posterior predictive distribution by drawing from $p(\hat{S}_{i^*j}(t) \mid \boldsymbol{x}_{i^*j}, \boldsymbol{z}_{i^*j}, \boldsymbol{\theta}^{(l)}, \boldsymbol{b}_j^{(l)})$ where $\boldsymbol{\theta}^{(l)}$ and $\boldsymbol{b}_j^{(l)}$ are the $l^{th}$ $(l = 1,...,L)$ MCMC draws from the joint posterior distribution $p(\boldsymbol{\theta}, \boldsymbol{b}_j \mid \mathcal{D})$. 

Alternatively, individual $i^*$ may come from a new cluster $j^* \neq j$ (for all $j \in \{1,...,J\}$) that was not in our sample. The predicted probability of being event-free at time $0 < t \leq T_{\max}$ is therefore denoted $\hat{S}_{i^*j^*}(t)$ and can be generated from the posterior predictive distribution:
\begin{equation}
\begin{aligned}
p \Big( \hat{S}_{i^*j^*}(t) \mid \boldsymbol{x}_{i^*j^*}, \boldsymbol{z}_{i^*j^*}, \mathcal{D} \Big) 
& = 
  \int
    \int
      p \Big( \hat{S}_{i^*j^*}(t) \mid \boldsymbol{x}_{i^*j^*}, \boldsymbol{z}_{i^*j^*}, \boldsymbol{\theta}, \boldsymbol{\tilde{b}}_{j^*} \Big)
      p \Big( \boldsymbol{\theta}, \boldsymbol{\tilde{b}}_{j^*} \mid \mathcal{D} \Big)
    d \boldsymbol{\tilde{b}}_{j^*} \space d \boldsymbol{\theta} \\
& = 
  \int
    \int
      p \Big( \hat{S}_{i^*j^*}(t) \mid \boldsymbol{x}_{i^*j^*}, \boldsymbol{z}_{i^*j^*}, \boldsymbol{\theta}, \boldsymbol{\tilde{b}}_{j^*} \Big)
      p \Big( \boldsymbol{\tilde{b}}_{j^*} \mid \boldsymbol{\theta} \Big)
      p \Big( \boldsymbol{\theta} \mid \mathcal{D} \Big)
    d \boldsymbol{\tilde{b}}_{j^*} \space d \boldsymbol{\theta} \\
\end{aligned}
\end{equation}
where $\boldsymbol{\tilde{b}}_{j^*}$ denotes the cluster-specific parameters for the new cluster. We can obtain draws for $\boldsymbol{\tilde{b}}_{j^*}$ during estimation of the model (in a similar manner as for $\boldsymbol{b}_j$). At the $l^{th}$ iteration of the MCMC sampler we obtain $\boldsymbol{\tilde{b}}_{j^*}^{(l)}$ as a random draw from the posterior distribution of the cluster-specific parameters and store it for later use in predictions. The set of random draws $\boldsymbol{\tilde{b}}_{j^*}^{(l)}$ for $l = 1,...,L$ then allow us to essentially marginalise over the distribution of the cluster-specific parameters. This is the approach used in \pkg{rstanarm} to generate survival predictions for individuals in new clusters that were not part of the original sample.

\subsection{Conditional survival probabilities}

In some instances we want to evaluate the predicted survival probability conditional on a last known survival time. This is known as a conditional survival probability.

Suppose that individual $i^*$ is known to be event-free up until $C_{i^*}$ and we wish to predict the survival probability at some time $t > C_{i^*}$. To do this we draw from the conditional posterior predictive distribution:
\begin{equation}
\begin{split}
p \Big( \hat{S}_{i^*}(t) \mid \boldsymbol{x}_{i^*}, \mathcal{D}, t > C_{i^*} \Big) = 
  \frac
    {p \Big( \hat{S}_{i^*}(t)      \mid \boldsymbol{x}_{i^*}, \mathcal{D} \Big)}
    {p \Big( \hat{S}_{i^*}(C_{i^*}) \mid \boldsymbol{x}_{i^*}, \mathcal{D} \Big)}
\end{split}
\end{equation}
or -- equivalently -- for multilevel survival models we have individual $i^*$ in cluster $j^*$ who is known to be event-free up until $C_{i^*j^*}$:
\begin{equation}
\begin{split}
p \Big( \hat{S}_{i^*j^*}(t) \mid \boldsymbol{x}_{i^*j^*}, \boldsymbol{z}_{i^*j^*}, \mathcal{D}, t > C_{i^*j^*} \Big) = 
  \frac
    {p \Big( \hat{S}_{i^*j^*}(t)        \mid \boldsymbol{x}_{i^*j^*}, \boldsymbol{z}_{i^*j^*}, \mathcal{D} \Big)}
    {p \Big( \hat{S}_{i^*j^*}(C_{i^*j^*}) \mid \boldsymbol{x}_{i^*j^*}, \boldsymbol{z}_{i^*j^*}, \mathcal{D} \Big)}
\end{split}
\end{equation}

\subsection{Standardised survival probabilities} \label{sec:standardised-survival}

All of the previously discussed predictions require conditioning on some covariate values $\boldsymbol{x}_{ij}$ and $\boldsymbol{z}_{ij}$. Even if we have a multilevel survival model and choose to marginalise over the distribution of the cluster-specific parameters, we are still obtaining predictions at some known unique values of the covariates.

However sometimes we wish to generate an "average" survival probability. One possible approach is to predict at the mean value of all covariates \citep{Cupples:1995}. However this doesn't always make sense, especially not in the presence of categorical covariates. For instance, suppose our covariates are gender and a treatment indicator. Then predicting for an individual at the mean of all covariates might correspond to a 50\% male who was 50\% treated. That does not make sense and is not what we wish to do.

A better alternative is to average over the individual survival probabilties. This essentially provides an approximation to marginalising over the joint distribution of the covariates. At any time $t$ it is possible to obtain a so-called standardised survival probability, denoted $\hat{S}^{*}(t)$, by averaging the individual-specific survival probabilities:
\begin{equation}
\begin{split}
p ( \hat{S}^{*}(t) \mid \mathcal{D} ) =
  \frac{1}{N^{P}}
    \sum_{i=1}^{N^{P}} p ( \hat{S}_i(t) \mid \boldsymbol{x}_{i^*}, \mathcal{D} )
\end{split}
\end{equation}
where $\hat{S}_i(t)$ is the predicted survival probability for individual $i$ ($i = 1,...,N^{P}$) at time $t$, and $N^{P}$ is the number of individuals included in the predictions. For multilevel survival models the calculation is similar and follows quite naturally (details not shown).

Note however that if $N^{P}$ is not sufficiently large (for example we predict individual survival probabilities using covariate data for just $N^{P} = 2$ individuals) then averaging over their covariate distribution may not be meaningful. Similarly, if we estimated a multilevel survival model and then predicted standardised survival probabilities based on just $N^{P} = 2$ individuals from our sample, the joint distribution of their cluster-specific parameters would likely be a poor representation of the distribution of cluster-specific parameters for the entire sample and population.

It is therefore better to calculate standardised survival probabilities by setting $N^{P}$ equal to the total number of individuals in the original sample (i.e. $N^{P} = N$. This approach can then also be used for assessing the fit of the survival model in \pkg{rstanarm} (see the \fct{ps_check} function described in Section \ref{sec:implementation}). Posterior predictive draws of the standardised survival probability are evaluated at a series of time points between 0 and $T_{\max}$ using all individuals in the estimation sample and the predicted standardised survival curve is overlaid with the observed Kaplan-Meier survival curve.


\section{Implementation} \label{sec:implementation}

\subsection{Overview}

The \pkg{rstanarm} package is built on top of the \pkg{rstan} \proglang{R} package \citep{Stan:2019}, which is the \proglang{R} interface for \proglang{Stan}. 
Models in \pkg{rstanarm} are written in the \proglang{Stan} programming language, translated into \proglang{C++} code, and then compiled at the time the package is built. This means that for most users -- who install a binary version of \pkg{rstanarm} from the Comprehensive R Archive Network (CRAN) -- the models in \pkg{rstanarm} will be pre-compiled. This is beneficial for users because there is no compilation time either during installation or when they estimate a model.

\subsection{Main modelling function}

Survival models in \pkg{rstanarm} are implemented around the \fct{stan_surv} modelling function.

The function signature for \fct{stan_surv} is:
\begin{Schunk}
\begin{Sinput}
R> stan_surv(formula, data, basehaz = "ms", basehaz_ops, qnodes = 15,
+    prior = normal(), prior_intercept = normal(), prior_aux,
+    prior_smooth = exponential(autoscale = FALSE),
+    prior_covariance = decov(), prior_PD = FALSE,
+    algorithm = c("sampling", "meanfield", "fullrank"),
+    adapt_delta = 0.95, ...)
\end{Sinput}
\end{Schunk}

The following provides a brief description of the main features of each of these arguments:

\begin{itemize}

\item The \code{formula} argument accepts objects built around the standard \proglang{R} formula syntax (see \fct{stats::formula}). The left hand side of the formula should be an object returned by the \fct{Surv} function in the \pkg{survival} package \citep{Therneau:2019}. Any random effects structure (for multilevel survival models) can be specified on the right hand side of the formula using the same syntax as the \pkg{lme4} \proglang{R} package \citep{Bates:2015} as shown in the example in Section \ref{sec:multilevelmodel}.

By default, any covariate effects specified in the fixed-effect part of the model \code{formula} are included under a proportional hazards assumption (for models estimated using a hazard scale formulation) or under the assumption of time-fixed acceleration factors (for models estimated using an AFT formulation). Time-varying effects are specified in the model \code{formula} by wrapping the covariate name in the \fct{tve} function. For example, if we wanted to estimate a time-varying effect for the covariate \code{sex} then we could specify \code{tve(sex)} in the model formula, e.g. \code{formula = Surv(time, status) ~ tve(sex) + age}. The \fct{tve} function is a special function that only has meaning when used in the \code{formula} of a model estimated using \fct{stan_surv}. Its functionality is demonstrated in the worked examples in Sections \ref{sec:tvebs} and \ref{sec:tvepw}.

\item The \code{data} argument accepts an object inheriting the class \class{data.frame}, in other words the usual \proglang{R} data frame.

\item The choice of parametric baseline hazard (or baseline survival distribution for AFT models) is specified via the \code{basehaz} argument. For the M-spline (\code{"ms"}) and B-spline (\code{"bs"}) models additional options related to the spline degree $\delta$, knot locations $\boldsymbol{k}$, or degrees of freedom $L$ can be specified as a list and passed to the \code{basehaz_ops} argument. For example, specifying \code{basehaz = "ms"} and \code{basehaz_ops = list(degree = 2, knots = c(10,20))} would request a baseline hazard modelled using quadratic M-splines with two internal knots located at $t = 10$ and $t = 20$.

\item The argument \code{qnodes} is a control argument that allows the user to specify the number of quadrature nodes when quadrature is required (as described in Section \ref{sec:loglikelihood}).

\item The \code{prior} family of arguments allow the user to specify the prior distributions for each of the parameters, as follows:

  \begin{itemize}
  
  \item \code{prior} relates to the time-fixed regression coefficients;
  
  \item \code{prior_intercept} relates to the intercept in the linear predictor;
  
  \item \code{prior_aux} relates to the so-called "auxiliary" parameters in the baseline hazard ($\gamma$ for the Weibull and Gompertz models or $\boldsymbol{\gamma}$ for the M-spline and B-spline models);
  
  \item \code{prior_smooth} relates to the hyperparameter $\tau_p$ when the $p^{th}$ covariate has a time-varying effect; and
  
  \item \code{prior_covariance} relates to the covariance matrix for the random effects when a multilevel survival model is being estimated.
  
  \end{itemize}

\item The remaining arguments (\code{prior_PD}, \code{algorithm}, and \code{adapt_delta}) are optional control arguments related to estimation in Stan:

  \begin{itemize}
  
  \item Setting \code{prior_PD = TRUE} states that the user only wants to draw from the prior predictive distribution and not condition on the data.
  
  \item The \code{algorithm} argument specifies the estimation routine to use. This includes either Hamiltonian Monte Carlo (\code{"sampling"}) or one of the variational Bayes algorithms (\code{"meanfield"} or \code{"fullrank"}). The model specification is agnostic to the chosen \code{algorithm}. That is, the user can choose from any of the available algorithms regardless of the specified model.
  
  \item The \code{adapt_delta} argument controls the target average acceptance probability. It is only relevant when \code{algorithm = "sampling"} in which case \code{adapt_delta} should be between 0 and 1, with higher values leading to smaller step sizes and therefore a more robust sampler but longer estimation times.
  
  \end{itemize}

\end{itemize}

The model returned by \fct{stan_surv} is an object of class \class{stansurv} and inheriting the \class{stanreg} class. It is effectively a list with a number of important attributes. There are a range of post-estimation functions that can be called on \class{stansurv} (and \class{stanreg}) objects -- some of the most important ones are described in Section \ref{sec:postest-functions}.

\subsubsection{Default knot locations}

Default knot locations for the M-spline, B-spline, or piecewise constant functions are the same regardless of whether they are used for modelling the baseline hazard or time-varying effects. By default the vector of knot locations $\boldsymbol{k} = \{k_{1},...,k_{J}\}$ includes a lower boundary knot $k_{1}$ at the earliest entry time (equal to zero if there isn't delayed entry) and an upper boundary knot $k_{J}$ at the latest event or censoring time. The location of the boundary knots cannot be changed by the user. 

Internal knot locations -- that is $k_{2},...,k_{(J-1)}$ when $J \geq 3$ -- can be explicitly specified by the user or are determined by default. The number of internal knots and/or their locations can be controlled via the \code{basehaz_ops} argument to \fct{stan_surv} (for modelling the baseline hazard) or via the arguments to the \fct{tve} function (for modelling a time-varying effect). If knot locations are not explicitly specified by the user, then the default is to place the internal knots at equally spaced percentiles of the distribution of uncensored event times. For instance, if there are three internal knots they would be placed at the $25^{\text{th}}$, $50^{\text{th}}$, and $75^{\text{th}}$ percentiles of the distribution of the uncensored event times. 

\subsection{Post-estimation functions} \label{sec:postest-functions}

The \pkg{rstanarm} package provides a range of post-estimation functions that can be used after fitting the survival model. This includes functions for inference (e.g. reporting parameter estimates), diagnostics (e.g. assessing model fit), and generating predictions. We highlight the most important ones here:

\begin{itemize}

\item The \fct{print} and \fct{summary} functions provide reports of parameter estimates and some summary information on the data (e.g. number of observations, number of events, etc). They each provide varying levels of detail. For example, the \fct{summary} method provides diagnostic measures such as Gelman and Rubin's Rhat statistic \citep{Gelman:1992} for assessing convergence of the MCMC chains and the number of effective MCMC samples. On the other hand, the \fct{print} method is more concise and does not provide this level of additional detail.

\item The \fct{fixef} and \fct{ranef} functions report the fixed effect and random effect parameter estimates, respectively.

\item The \fct{posterior_survfit} function is the primary function for generating survival predictions. The type of prediction is specified via the \code{type} arguments and can currently be any of the following:

  \begin{itemize}
  
  \item \code{"surv"}: the estimated survival probability;
  \item \code{"cumhaz"}: the estimated cumulative hazard;
  \item \code{"haz"}: the estimated hazard rate;
  \item \code{"cdf"}: the estimated failure probability;
  \item \code{"logsurv"}: the estimated log survival probability;
  \item \code{"logcumhaz"}: the estimated log cumulative hazard;
  \item \code{"loghaz"}: the estimated log hazard rate; or
  \item \code{"logcdf"}: the estimated log failure probability.
  
  \end{itemize}

There are additional arguments to \fct{posterior_survfit} that control the time at which the predictions are generated (\code{times}), whether they are generated across a time range (referred to as extrapolation, see \code{extrapolate}), whether they are conditional on a last known survival time (\code{condition}), and whether they are averaged across individuals (referred to as standardised predictions, see \code{standardise}). The returned predictions are a data frame with a special class called \class{survfit.stansurv}. The \class{survfit.stansurv} class has both \fct{print} and \fct{plot} methods that can be called on it. These will be demonstrated as part of the examples in Section \ref{sec:usage}.

\item The \fct{loo} and \fct{waic} functions report model fit statistics. The former is based on approximate leave-one-out cross validation \citep{Vehtari:2017} and is recommended. The latter is a less preferable alternative that reports the Widely Applicable Information Criterion (WAIC) criterion \citep{Watanabe:2010}. Both of these functions are built on top of the \pkg{loo} \proglang{R} package \citep{Vehtari:2019}. The values (objects) returned by either \fct{loo} or \fct{waic} can also be passed to the \fct{loo_compare} function to compare different models estimated on the same dataset. This will be demonstrated as part of the examples in Section  \ref{sec:usage}.

\item The \fct{log_lik} function generates a pointwise log likelihood matrix. That is, it calculates the log likelihood for each observation (either in the original dataset or some new dataset) using each MCMC draw of the model parameters.

\item The \fct{plot} function allows for a variety of plots depending on the input to the \code{plotfun} argument. The default is to plot the estimated baseline hazard (\code{plotfun = "basehaz"}), but alternatives include a plot of the estimated time-varying hazard ratio for models with time-varying effects (\code{plotfun = "tve"}), plots summarising the parameter estimates (e.g. posterior densities or posterior intervals), and plots providing diagnostics (e.g. MCMC trace plots).

\item The \fct{ps_check} function provides a quick diagnostic check for the fitted survival function. It is based on the estimation sample and compares the predicted standardised survival curve to the observed Kaplan-Meier survival curve.

\end{itemize}


\section{Usage examples} \label{sec:usage}

\subsection{A flexible parametric proportional hazards model}

In this example we fit a proportional hazards model with a flexible parametric baseline hazard modelled using cubic M-splines. This is an example of the default survival model estimated by the \fct{stan_surv} function in \pkg{rstanarm}.

We will use the German Breast Cancer Study Group dataset \citep{Royston:2002}. The data consist of $N = 686$ patients with primary node positive breast cancer recruited between 1984-1989. The primary response is time to recurrence or death. Median follow-up time was 1084 days. Overall, there were 299 (44\%) events and the remaining 387 (56\%) individuals were right censored. We concern our analysis here with a 3-category baseline covariate for cancer prognosis (good/medium/poor).

Let us fit the proportional hazards model:
\begin{Schunk}
\begin{Sinput}
R> mod1 <- stan_surv(formula = Surv(recyrs, status) ~ group, 
+                    data    = bcancer, 
+                    chains  = CHAINS, 
+                    cores   = CORES, 
+                    seed    = SEED,
+                    iter    = ITER)
\end{Sinput}
\end{Schunk}

Since there are no time-varying effects in the model (i.e. we did not wrap any covariates in the \fct{tve} function) there is a closed form expression for the cumulative hazard and survival function and so the model is relatively fast to fit. Specifically, the model takes ~3.5 seconds for each MCMC chain based on the default 2000 MCMC iterations (1000 warm up, 1000 sampling) on a standard desktop.

We can easily obtain the estimated hazard ratios for the 3-category group covariate using the generic \fct{print} method for \class{stansurv} objects, as follows:
\begin{Schunk}
\begin{Sinput}
R> print(mod1, digits = 2)
\end{Sinput}
\begin{Soutput}
stan_surv
 baseline hazard: M-splines on hazard scale
 formula:         Surv(recyrs, status) ~ group
 observations:    686
 events:          299 (43.6%)
 right censored:  387 (56.4%)
 delayed entry:   no
------
                Median MAD_SD exp(Median)
(Intercept)     -0.65   0.18     NA      
groupMedium      0.82   0.17   2.28      
groupPoor        1.60   0.15   4.95      
m-splines-coef1  0.00   0.00     NA      
m-splines-coef2  0.02   0.01     NA      
m-splines-coef3  0.40   0.07     NA      
m-splines-coef4  0.06   0.05     NA      
m-splines-coef5  0.21   0.12     NA      
m-splines-coef6  0.30   0.16     NA      

------
* For help interpreting the printed output see ?print.stanreg
* For info on the priors used see ?prior_summary.stanreg
\end{Soutput}
\end{Schunk}

From this output we see that individuals in the groups with \code{Poor} or \code{Medium} prognosis have much higher rates of death relative to the group with \code{Good} prognosis. The hazard of death in the \code{Poor} prognosis group is approximately 5-fold higher than the hazard of death in the \code{Good} prognosis group. Similarly, the hazard of death in the \code{Medium} prognosis group is approximately 2-fold higher than the hazard of death in the \code{Good} prognosis group.

It may also be of interest to compare the different types of parametric baseline hazards we could have used for this model (some more restricting, others more flexible). Let us fit several models, each with a different baseline hazard:
\begin{Schunk}
\begin{Sinput}
R> mod1_exp      <- update(mod1, basehaz = "exp")
R> mod1_weibull  <- update(mod1, basehaz = "weibull")
R> mod1_gompertz <- update(mod1, basehaz = "gompertz")
R> mod1_bspline  <- update(mod1, basehaz = "bs")
R> mod1_mspline1 <- update(mod1, basehaz = "ms")
R> mod1_mspline2 <- update(mod1, basehaz = "ms", basehaz_ops = list(df = 9))
\end{Sinput}
\end{Schunk}

The default action of the \fct{plot} method for \class{stansurv} objects is to plot the estimated baseline hazard. We will use this method to plot the baseline hazard for each of the competing models. First, we write a little helper function to adjust the y-axis limits and add a centered title on each plot, as follows:
\begin{Schunk}
\begin{Sinput}
R> plotfun <- function(model, title) {
+    plot(model, plotfun = "basehaz") +              
+      coord_cartesian(ylim = c(0,0.4)) +           
+      labs(title = title) +                        
+      theme(plot.title = element_text(hjust = 0.5)) 
+  }
\end{Sinput}
\end{Schunk}
and then we generate each of the plots:
\begin{Schunk}
\begin{Sinput}
R> p_exp      <- plotfun(mod1_exp,      "Exponential")
R> p_weibull  <- plotfun(mod1_weibull,  "Weibull")
R> p_gompertz <- plotfun(mod1_gompertz, "Gompertz")
R> p_bspline  <- plotfun(mod1_bspline,  "B-splines with\ntwo internal knots")
R> p_mspline1 <- plotfun(mod1_mspline1, "M-splines with\ntwo internal knots")
R> p_mspline2 <- plotfun(mod1_mspline2, "M-splines with\nfive internal knots")
\end{Sinput}
\end{Schunk}
and then combine the plots using the \fct{plot_grid} function from the \pkg{cowplot} \proglang{R} package \citep{Wilke:2019}:
\begin{Schunk}
\begin{Sinput}
R> p_combined <- plot_grid(p_exp,
+                          p_weibull,
+                          p_gompertz,
+                          p_bspline,
+                          p_mspline1,
+                          p_mspline2,
+                          ncol = 3)
\end{Sinput}
\end{Schunk}

\begin{figure}[t!]
\centering
\includegraphics{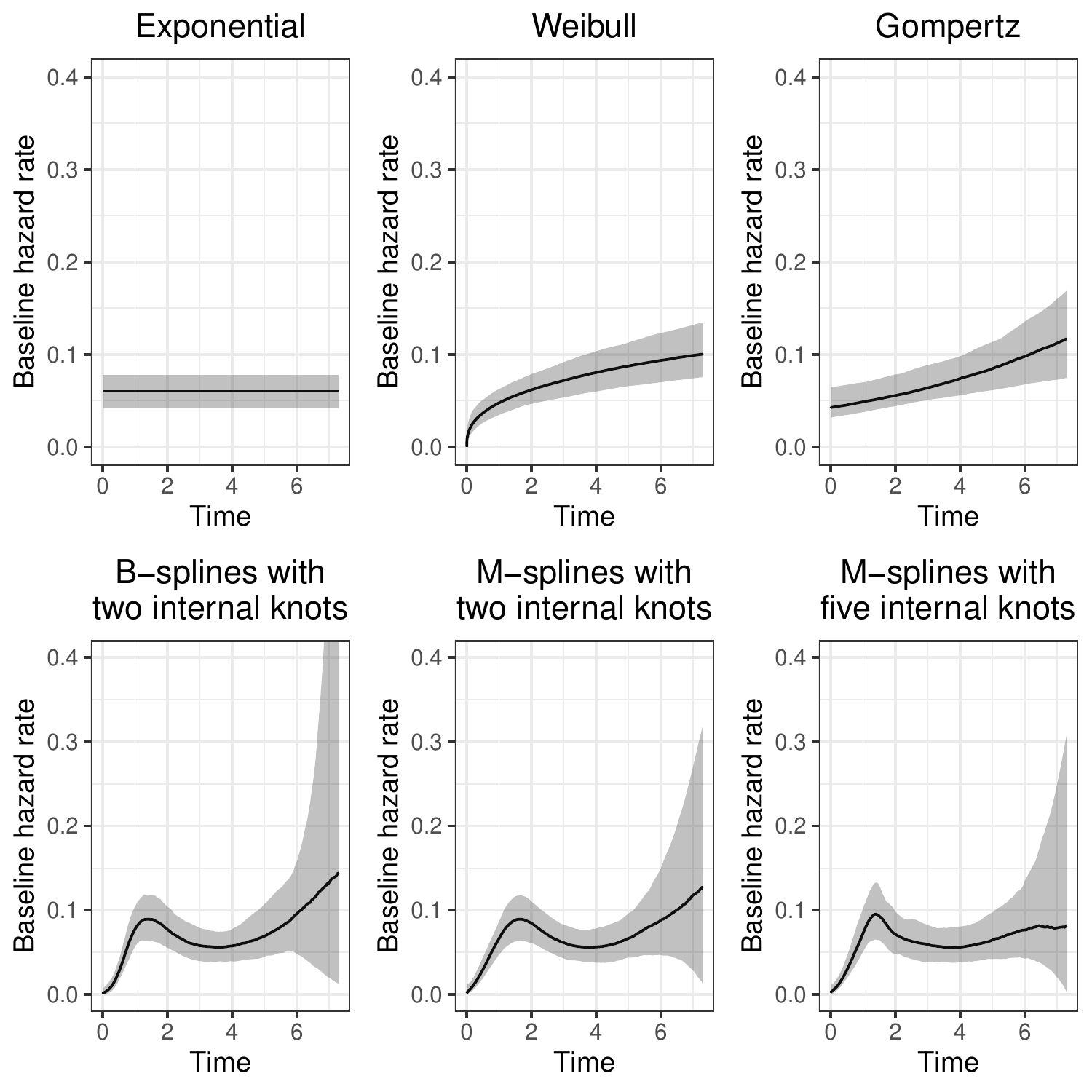}
\caption{\label{fig:basehaz-compare} Estimated baseline hazards (posterior median and 95\% uncertainty limits) for each of the six different models.}
\end{figure}

Figure \ref{fig:basehaz-compare} shows the resulting plot with the estimated baseline hazard for each model and 95\% posterior uncertainty limits. We can clearly see from the plot the additional flexibility the cubic spline models provide. They are able to capture at least two turning points in the hazard function, one around 1.5 years and another one around 4 years. It is also interesting to note that there appears to be very little change in the fit of the M-spline model when we increase the number of internal knots from two to five.

We can also compare the fit of these models using the \code{loo} method for \code{stansurv} objects:
\begin{Schunk}
\begin{Sinput}
R> compare_models(loo(mod1_exp),
+                 loo(mod1_weibull),
+                 loo(mod1_gompertz),
+                 loo(mod1_bspline),
+                 loo(mod1_mspline1),
+                 loo(mod1_mspline2))
\end{Sinput}
\begin{Soutput}
Model formulas: 
 :  NULL
 :  NULL
 :  NULL
 :  NULL
 :  NULL
 :  NULL              elpd_diff se_diff
mod1_mspline1   0.0       0.0  
mod1_bspline   -0.4       1.7  
mod1_mspline2  -1.6       1.5  
mod1_weibull  -18.0       5.3  
mod1_gompertz -31.5       6.1  
mod1_exp      -36.3       6.0  
\end{Soutput}
\end{Schunk}
where we see that models with a flexible parametric (spline-based) baseline hazard fit the data best followed by the standard parametric (Weibull, Gompertz, exponential) models. Roughly speaking, the B-spline and M-spline models seem to fit the data equally well since the differences in \code{elpd} between the models are very small relative to their standard errors. Moreover, increasing the number of internal knots for the M-splines from two (the default) to five doesn't seem to improve the fit (that is the default number of knots seems to provide sufficient flexibility for modelling the baseline hazard).

After fitting the survival model we often want to estimate the predicted survival function for individuals with different covariate patterns. Therefore, let us obtain the predicted survival function between 0 and 5 years for an individual in each of the prognostic groups. To do this we can use the \fct{posterior_survfit} method for \class{stansurv} objects and it's associated \fct{plot} method.

First let us construct the prediction (covariate) data:
\begin{Schunk}
\begin{Sinput}
R> nd <- data.frame(group = c("Good", "Medium", "Poor"))
R> head(nd)
\end{Sinput}
\begin{Soutput}
   group
1   Good
2 Medium
3   Poor
\end{Soutput}
\end{Schunk}
and then generate the posterior predictions:
\begin{Schunk}
\begin{Sinput}
R> ps <- posterior_survfit(mod1, 
+                          newdata     = nd, 
+                          times       = 0, 
+                          extrapolate = TRUE,
+                          control     = list(edist = 5))
R> head(ps)
\end{Sinput}
\begin{Soutput}
stan_surv predictions
 num. individuals: 3 
 prediction type:  event free probability 
 standardised?:    no 
 conditional?:     no 

  id cond_time   time median  ci_lb  ci_ub
1  1        NA 0.0000 1.0000 1.0000 1.0000
2  1        NA 0.0505 0.9999 0.9993 1.0000
3  1        NA 0.1010 0.9996 0.9987 0.9999
4  1        NA 0.1515 0.9992 0.9979 0.9997
5  1        NA 0.2020 0.9987 0.9971 0.9995
6  1        NA 0.2525 0.9981 0.9960 0.9991
\end{Soutput}
\end{Schunk}

Here we note that the \code{id} variable in the data frame of posterior predictions identifies which row of \code{newdata} the predictions correspond to. For demonstration purposes we have also included other arguments in the \fct{posterior_survfit} call, namely:

\begin{itemize}

\item the \code{times = 0} argument says that we want to predict at time $t = 0$ (i.e. baseline) for each individual in the \code{newdata} (this is the default anyway);

\item the \code{extrapolate = TRUE} argument says that we want to extrapolate forward from time $t = 0$ (this is also the default); and

\item the \code{control = list(edist = 5)} identifies the control of the extrapolation; this is saying we wish to extrapolate the survival function forward from time $t = 0$ for a distance of 5 time units (the default would have been to extrapolate as far as the largest event or censoring time in the estimation dataset, which is 7.28 years in the \code{bcancer} data).

\end{itemize}

Let us now plot the survival predictions. We will relabel the \code{id} variable with meaningful labels identifying the covariate profile of each new individual in our prediction data:
\begin{Schunk}
\begin{Sinput}
R> panel_labels <- c('1' = "Good",
+                    '2' = "Medium",
+                    '3' = "Poor")
R> pps <- plot(ps) + 
+    facet_wrap(~ id, labeller = labeller(id = panel_labels))
\end{Sinput}
\end{Schunk}

\begin{figure}[t!]
\centering
\includegraphics{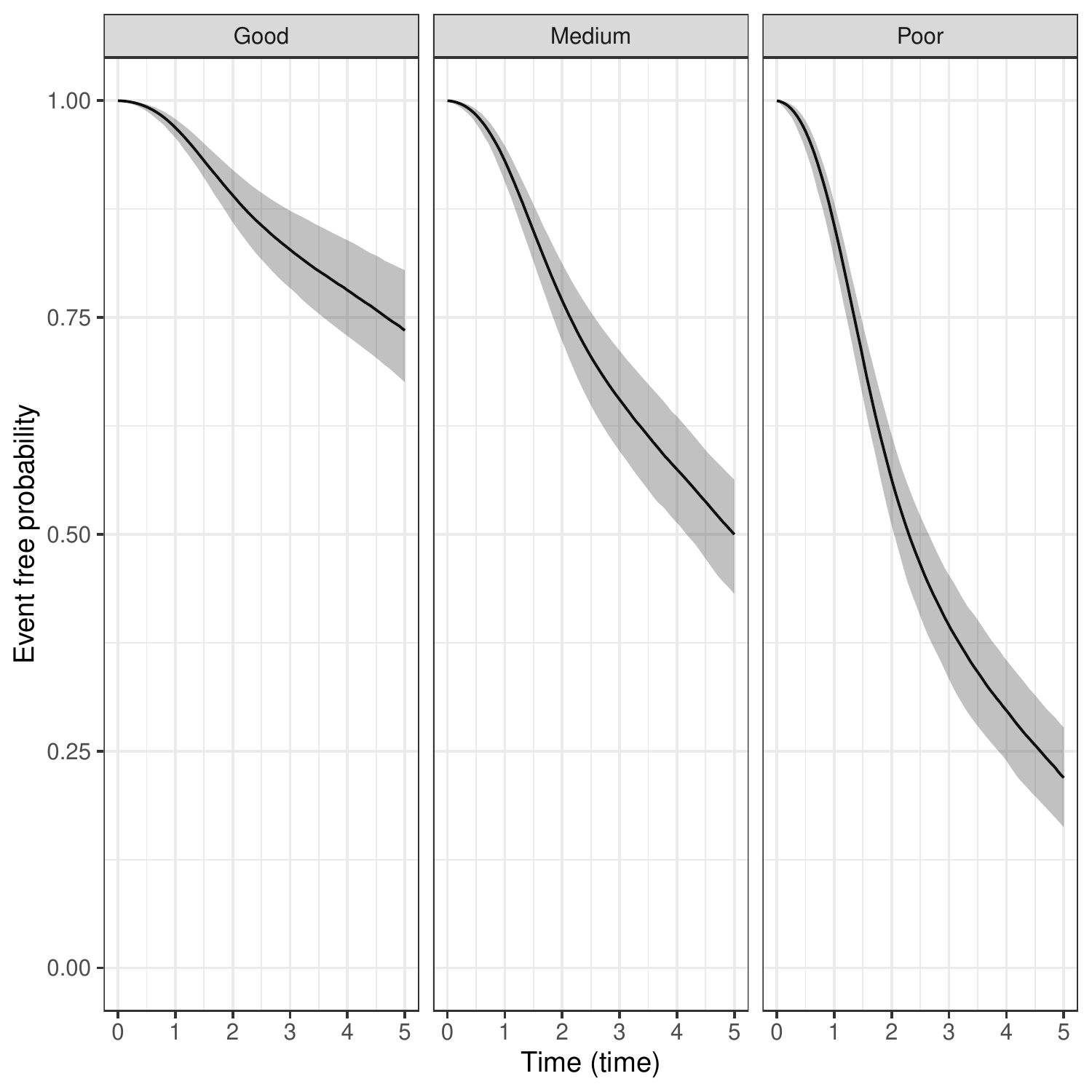}
\caption{\label{fig:predsurv} Predicted survival function (posterior median and 95\% uncertainty limits) for an individual in either the good, medium, or poor prognosis group.}
\end{figure}

Figure \ref{fig:predsurv} shows the resulting plot. We can see from the plot that predicted survival is worst for patients with a \code{Poor} diagnosis, and best for patients with a \code{Good} diagnosis, as we would expect based on our previous model estimates.

Alternatively, if we wanted to obtain the predicted hazard or log hazard function for each individual in our new data (instead of their survival function), then we just need to specify \code{type = "haz"} or \code{"loghaz"} in our \fct{posterior_survfit} call (the default is \code{type = "surv"}), as follows:
\begin{Schunk}
\begin{Sinput}
R> ph <- posterior_survfit(mod1, newdata = nd, type = "haz")
R> pl <- posterior_survfit(mod1, newdata = nd, type = "loghaz")
\end{Sinput}
\end{Schunk}
and then we can plot the predicted hazard:
\begin{Schunk}
\begin{Sinput}
R> pph <- plot(ph) + 
+    facet_wrap(~ id, labeller = labeller(id = panel_labels))
R> ppl <- plot(pl) + 
+    facet_wrap(~ id, labeller = labeller(id = panel_labels))
\end{Sinput}
\end{Schunk}

\begin{figure}[t!]
\centering
\includegraphics{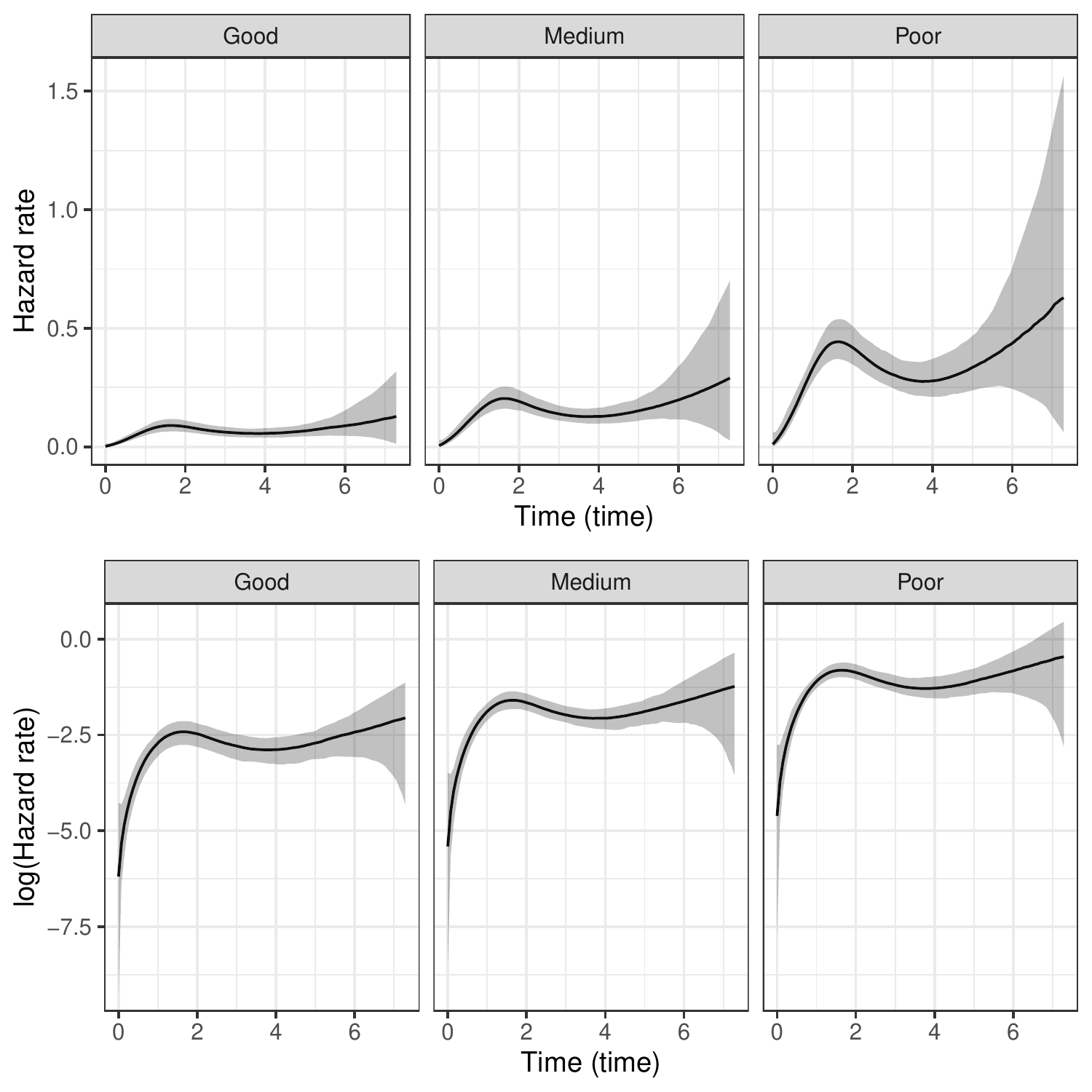}
\caption{\label{fig:predhaz} Predicted hazard function (top row) and log hazard function (bottom row) (posterior median and 95\% uncertainty limits) for an individual in either the good, medium, or poor prognosis group.}
\end{figure}

Figure \ref{fig:predhaz} shows the resulting plot. We can quite clearly see in the plot the assumption of proportional hazards, i.e. proportional lines for the hazards and parallel lines for the log hazards. We can also see that the hazard is highest in the \code{Poor} prognosis group (i.e. worst survival) and the hazard is lowest in the \code{Good} prognosis group (i.e. best survival). This corresponds to relationships we saw in the plot of the survival functions in Figure \ref{fig:predsurv}.

\subsection{A standard parametric AFT model} \label{sec:aftmodel}

In this example we demonstrate the relationship between a Weibull AFT model and its proportional hazards reparameterisation. We assume a Weibull distribution for the event times since the Weibull distribution has both a proportional hazards and an AFT parameterisation as was described in Section \ref{sec:modelformulations}.

We will again use the German Breast Cancer Study Group dataset that was introduced in the previous example.

Let us first fit the Weibull AFT model for time to recurrence or death with our 3-category baseline covariate for cancer prognosis:
\begin{Schunk}
\begin{Sinput}
R> m_aft <- stan_surv(formula = Surv(recyrs, status) ~ group, 
+                     data    = bcancer,
+                     basehaz = "weibull-aft", 
+                     chains  = CHAINS, 
+                     cores   = CORES, 
+                     seed    = SEED,
+                     iter    = ITER)
\end{Sinput}
\end{Schunk}
and then extract the estimated survival time ratios for cancer prognosis:
\begin{Schunk}
\begin{Sinput}
R> tr <- exp(fixef(m_aft))[c('groupMedium', 'groupPoor')]
R> print(tr)
\end{Sinput}
\begin{Soutput}
groupMedium   groupPoor 
  0.5442187   0.2992096 
\end{Soutput}
\end{Schunk}

We can then fit an otherwise equivalent Weibull proportional hazards model. All that we need to do is specify \code{"weibull"} instead of \code{"weibull-aft"} in the \code{basehaz} argument:
\begin{Schunk}
\begin{Sinput}
R> m_ph <- update(m_aft, basehaz = "weibull")
\end{Sinput}
\end{Schunk}
and then extract the estimated hazard ratios for cancer prognosis:
\begin{Schunk}
\begin{Sinput}
R> hr <- exp(fixef(m_ph))[c('groupMedium', 'groupPoor')]
R> print(hr)
\end{Sinput}
\begin{Soutput}
groupMedium   groupPoor 
   2.356028    5.310558 
\end{Soutput}
\end{Schunk}

We then use the relationship described in Section \ref{sec:aft-haz-relationship} to convert the estimated coefficients from the AFT model (i.e. log survival time ratios) to log hazard ratios. This can be done as follows:
\begin{Schunk}
\begin{Sinput}
R> aft_betas <- fixef(m_aft)[c("groupMedium", "groupPoor")]
R> aft_gamma <- fixef(m_aft)[c("weibull-shape")]
R> aft_hr    <- exp(- aft_gamma * aft_betas)
\end{Sinput}
\end{Schunk}

We can now compare the hazard ratios derived from the AFT parameterisation with the hazard ratios estimated using the proportional hazards parameterisation:
\begin{Schunk}
\begin{Sinput}
R> cbind("HR (PH model)"  = hr,
+        "HR (AFT model)" = aft_hr)
\end{Sinput}
\begin{Soutput}
            HR (PH model) HR (AFT model)
groupMedium      2.356028       2.303716
groupPoor        5.310558       5.233392
\end{Soutput}
\end{Schunk}

They agree closely, with slight differences due to sampling variation.

We can also look at the mean of the log posterior of the models:
\begin{Schunk}
\begin{Sinput}
R> cbind("LP (PH model)"  = summary(m_ph,  par = "log-posterior")[, "mean"],
+        "LP (AFT model)" = summary(m_aft, par = "log-posterior")[, "mean"])
\end{Sinput}
\begin{Soutput}
     LP (PH model) LP (AFT model)
[1,]     -821.3624       -821.173
\end{Soutput}
\end{Schunk}

We see that they are effectively equal, with slight differences due to sampling variation. This further demonstrates that these are in fact two different parameterisations of the same Weibull model.

\subsection{Time-varying covariates} \label{sec:tvc}

We demonstrate estimation of a model with time-varying covariates using the \code{pbc} data from the \pkg{survival} package. The data contains survival information for 312 patients with primary biliary cirrhosis who participated in a randomised placebo controlled trial of D-penicillamine conducted at the Mayo Clinic between 1974 and 1984.

The \pkg{rstanarm} package contains a small subset of patients ($N = 40$) from the PBC trial. However in this example we use data from all 312 patients so we will load the dataset from the \pkg{survival} package instead. There are in fact two datasets with relevant information:

\begin{itemize}

\item \code{pbc}: contains survival and transplant information for each patient (one row per patient); and

\item \code{pbcseq}: contains longitudinal biomarker measurements (multiple rows per patient).

\end{itemize}

We use functionality from the \pkg{survival} package to merge the two datasets so that they form a single long format dataset with a so-called "start/stop" structure. For comparison purposes we follow the same approach for constructing the data and fitting our model as described in the "Using Time Dependent Covariates" vignette for the \pkg{survival} \proglang{R} package \citep{Therneau:2019}:

\begin{Schunk}
\begin{Sinput}
R> dat <- survival::pbc
R> dat <- dat[dat$id <= 312, ]
R> dat <- tmerge(dat, dat, id = id, 
+                death = event(time, as.numeric(status == 2)))
R> dat <- tmerge(dat, survival::pbcseq, id = id, 
+                ascites  = tdc(day, ascites),
+                bili     = tdc(day, bili), 
+                albumin  = tdc(day, albumin),
+                protime  = tdc(day, protime), 
+                alk.phos = tdc(day, alk.phos))
R> dat <- dat[, c("id", "tstart", "tstop", "death", "bili", "protime")]
R> head(dat, 11)
\end{Sinput}
\begin{Soutput}
   id tstart tstop death bili protime
1   1      0   192     0 14.5    12.2
2   1    192   400     1 21.3    11.2
3   2      0   182     0  1.1    10.6
4   2    182   365     0  0.8    11.0
5   2    365   768     0  1.0    11.6
6   2    768  1790     0  1.9    10.6
7   2   1790  2151     0  2.6    11.3
8   2   2151  2515     0  3.6    11.5
9   2   2515  2882     0  4.2    11.5
10  2   2882  3226     0  3.6    11.5
11  2   3226  4500     0  4.6    11.5
\end{Soutput}
\end{Schunk}

The output shows the first 11 rows of the resulting dataset, including the "start-stop" structure. The \code{tstart} and \code{tstop} variables denote a time interval. The longitudinal biomarker measurements (\code{bili} and \code{protime}) are assumed to remain constant within the time interval \code{[tstart,tstop)} and the event indicator (\code{death}) is assumed to occur at the end of the interval, i.e. when $t$ is equal to \code{tstop}.

We use this data to investigate the association between the hazard of death and two longitudinal biomarkers: bilirubin and prothrombin. Higher values of bilirubin or prothrombin are associated with worse liver function. Therefore in patients with primary biliary cirrhosis we expect that higher bilirubin or prothrombin values should be associated with a higher hazard of death. 

We estimate our model as follows:

\begin{Schunk}
\begin{Sinput}
R> mod_tvc <- stan_surv(
+    formula = Surv(tstart, tstop, death) ~ log(bili) + log(protime), 
+    data    = dat,
+    chains  = CHAINS, 
+    cores   = CORES, 
+    seed    = SEED,
+    iter    = ITER)
\end{Sinput}
\end{Schunk}

and examine our estimated parameters as usual:

\begin{Schunk}
\begin{Sinput}
R> print(mod_tvc, digits = 2)
\end{Sinput}
\begin{Soutput}
stan_surv
 baseline hazard: M-splines on hazard scale
 formula:         Surv(tstart, tstop, death) ~ log(bili) + log(protime)
 observations:    1807
 events:          125 (6.9%)
 right censored:  1682 (93.1%)
 delayed entry:   yes
------
                Median MAD_SD exp(Median)
(Intercept)     -11.97   1.04     NA     
log(bili)         1.28   0.09   3.58     
log(protime)      4.24   0.40  69.42     
m-splines-coef1   0.04   0.02     NA     
m-splines-coef2   0.05   0.03     NA     
m-splines-coef3   0.21   0.07     NA     
m-splines-coef4   0.21   0.13     NA     
m-splines-coef5   0.30   0.16     NA     
m-splines-coef6   0.17   0.13     NA     

------
* For help interpreting the printed output see ?print.stanreg
* For info on the priors used see ?prior_summary.stanreg
\end{Soutput}
\end{Schunk}

Here we strong evidence that higher log bilirubin or log prothrombin are associated with a higher hazard of death. However there are some other aspects worth noting in the output. First, the reported number of observations is 1807. This is not the same as the number of patients in our dataset ($N = 312$). Second, the reported number of right censored observations in 1682. This is also much greater than the number of individuals in our dataset.

These discrepancies occur because we transformed our data into a "start-stop" structure with multiple rows for each patient. When estimating the model, \fct{stan_surv} treats each row of the transformed data as a separate observation. For instance, recall that patient \code{id = 1} had two rows of data. The first row of their data represented the time interval \code{[0,192]} at the end of which they were right censored (i.e. still alive). The second row of their data represented the time interval \code{[192,400]} at the end of which they died. Therefore patient \code{id = 1} contributes these two rows as separate observations to the log likelihood of the model. Moreover, to accommodate their second row of data we have to be able to allow for delayed entry because the time interval over which they were observed was $[192,400)$. That is, their second row of data did not start from time 0. This demonstrates that the handling of time-varying covariates relies on the ability to handle delayed entry (i.e. they both rely on a "start-stop" data structure).

This highlights one important consideration with \fct{stan_surv}. Namely, that the definition of an observation in \fct{stan_surv} is a row of data and not specifically an individual. This may be of importance when comparing models using approximate leave-one-out cross validation.

For example, the default \fct{loo} method can be called as follows:

\begin{Schunk}
\begin{Sinput}
R> loo(mod_tvc)
\end{Sinput}
\begin{Soutput}
Computed from 500 by 1807 log-likelihood matrix

         Estimate    SE
elpd_loo  -1019.5  80.8
p_loo         9.0   2.2
looic      2039.1 161.7
------
Monte Carlo SE of elpd_loo is NA.

Pareto k diagnostic values:
                         Count Pct.    Min. n_eff
(-Inf, 0.5]   (good)     1806  99.9%   204       
 (0.5, 0.7]   (ok)          0   0.0%   <NA>      
   (0.7, 1]   (bad)         1   0.1%   42        
   (1, Inf)   (very bad)    0   0.0%   <NA>      
See help('pareto-k-diagnostic') for details.
\end{Soutput}
\end{Schunk}

However the default calculation assumes that we wish to "leave out" one row of data. But perhaps it would make more sense to evaluate \fct{loo} by "leaving out" an individual rather than an observation. To achieve this we must do the following. First we must generate the pointwise log likelihood matrix (i.e. the log likelihood for observation at each MCMC draw) using the \fct{log_lik} function, then collapse (i.e. sum) the log likelihood within an individual, and then pass the resulting matrix to the \fct{loo} function. This can be achieved as follows:

\begin{Schunk}
\begin{Sinput}
R> ids <- dat$id
R> ll  <- log_lik(mod_tvc)
R> ll  <- apply(ll, 1L, function(row) tapply(row, ids, sum))
R> ll  <- t(ll)
R> loo(ll)
\end{Sinput}
\begin{Soutput}
Computed from 500 by 312 log-likelihood matrix

         Estimate    SE
elpd_loo  -1019.5  67.8
p_loo         8.9   2.5
looic      2039.0 135.5
------
Monte Carlo SE of elpd_loo is 0.2.

All Pareto k estimates are good (k < 0.5).
See help('pareto-k-diagnostic') for details.
\end{Soutput}
\end{Schunk}

\subsection{Non-proportional hazards modelled using B-splines} \label{sec:tvebs}

To demonstrate the implementation of time-varying effects in \fct{stan_surv} we will use a simulated dataset, generated using the \pkg{simsurv} \proglang{R} package \citep{Brilleman:2019}.

We will simulate a dataset with $N = 500$ individuals with event times generated under the following Weibull hazard function:
\begin{equation}
h_i(t) = \gamma t^{\gamma-1} \lambda \exp( \beta(t) x_i )
\end{equation}
with scale parameter $\lambda = 0.1$, shape parameter $\gamma = 1.5$, binary baseline covariate $X_i \sim \text{Bern}(0.5)$, and time-varying hazard ratio $\beta(t) = -0.5 + 0.2 t$. We will enforce administrative censoring at 5 years if an individual's simulated event time is >5 years. In the code below \code{N} is used to represent the number of individuals:

\begin{Schunk}
\begin{Sinput}
R> set.seed(999111)
R> N <- 500
R> covs <- data.frame(id  = 1:N, 
+                     trt = rbinom(N, 1L, 0.5))
R> dat  <- simsurv(dist    = "weibull",
+                  lambdas = 0.1, 
+                  gammas  = 1.5, 
+                  betas   = c(trt = -0.5),
+                  tde     = c(trt = 0.2),
+                  x       = covs, 
+                  maxt    = 5)
R> dat  <- merge(dat, covs)
R> head(dat)
\end{Sinput}
\begin{Soutput}
  id eventtime status trt
1  1 2.5099804      1   0
2  2 4.8693271      1   1
3  3 3.3246030      1   1
4  4 0.3595983      1   0
5  5 0.6424857      1   1
6  6 1.4652469      1   1
\end{Soutput}
\end{Schunk}

With this simulated dataset we fit a model with a Weibull baseline hazard and a time-varying hazard ratio for \code{trt}:
\begin{Schunk}
\begin{Sinput}
R> mod2 <- stan_surv(
+    formula = Surv(eventtime, status) ~ tve(trt), 
+    data    = dat,
+    basehaz = "weibull",
+    chains  = CHAINS, 
+    cores   = CORES, 
+    seed    = SEED,
+    iter    = ITER)
\end{Sinput}
\end{Schunk}

The \fct{tve} function is used in the model formula to state that we want a time-varying effect (i.e. a time-varying coefficient) to be estimated for the variable \code{trt}. By default, a cubic B-spline basis with 3 degrees of freedom (i.e. two boundary knots placed at the limits of the range of event times, but no internal knots) is used for modelling the time-varying log hazard ratio. If we wanted to change the degree, knot locations, or degrees of freedom for the B-spline function we can specify additional arguments to the \fct{tve} function.

For example, to model the time-varying log hazard ratio using quadratic B-splines with 4 degrees of freedom (i.e. two boundary knots placed at the limits of the range of event times, as well as two internal knots placed -- by default -- at the 33.3rd and 66.6th percentiles of the distribution of uncensored event times) we could specify the model formula as:

\begin{Schunk}
\begin{Sinput}
R> Surv(eventtime, status) ~ tve(trt, df = 4, degree = 2)
\end{Sinput}
\end{Schunk}

\begin{figure}[t!]
\centering
\includegraphics{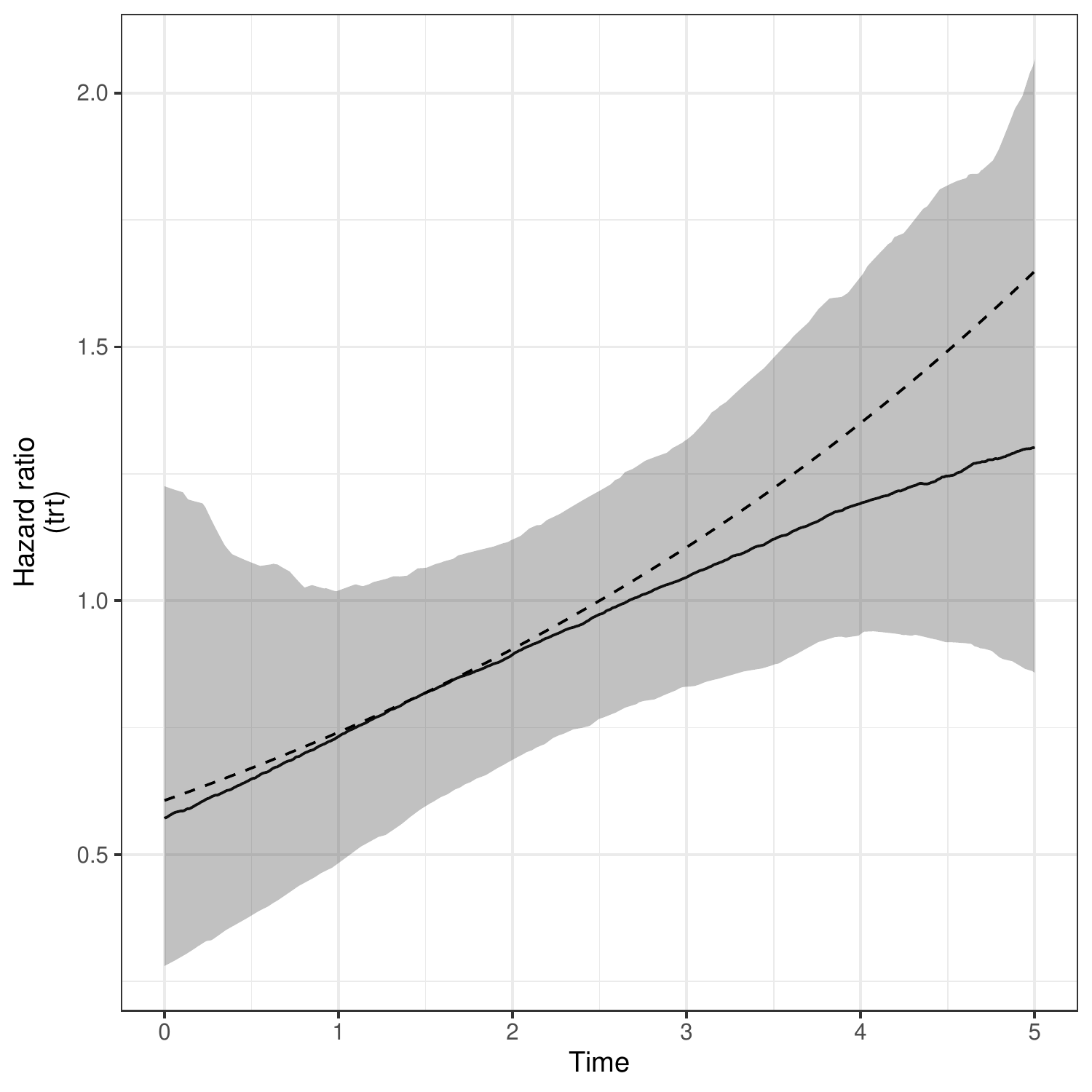}
\caption{\label{fig:tve-hr-bs} Time-varying hazard ratio for the estimated treatment effect (posterior median and 95\% uncertainty limits) when modelled using a smooth cubic B-spline function. The dashed line shows the "true" time-varying hazard ratio used to simulate the data.}
\end{figure}

Figure \ref{fig:tve-hr-bs} shows the estimated time-varying hazard ratio from the fitted model. This figure was obtained using the generic \fct{plot} method for \class{stansurv} objects and specifying the \code{plotfun = "tve"} argument (noting that in this case there is only one covariate in the model with a time-varying effect, but if there were others we could specify which covariate(s) we want to plot the time-varying effect for by specifying the \code{pars} argument to the \fct{plot} method).

From Figure \ref{fig:tve-hr-bs} we can see how the hazard ratio (i.e. the effect of treatment on the hazard of the event) changes as a function of time. The treatment appears to be protective during the first few years following baseline (i.e. HR < 1), and then the treatment appears to become harmful after about 2.5 years post-baseline. This is a reflection of the model we simulated under. 

Figure \ref{fig:tve-hr-bs} also shows a large amount of uncertainty around the estimated time-varying hazard ratio. This is to be expected, since we simulated a dataset of 500 individuals of which only around 70\% experienced the event before being censored at 5 years. So there are relatively few events with which to reliably estimate the time-varying hazard ratio. In general, we require a much larger number of events in our data in order to estimate a time-varying effect reliably when compared with a time-fixed effect. This is because a time-fixed effect essentially uses information about event rates that can be averaged across the entire time range.

\subsection{Non-proportional hazards modelled using a piecewise constant function} \label{sec:tvepw}

In the previous example we showed how non-proportional hazards can be modelled by using a smooth cubic B-spline function for the time-varying log hazard ratio. This is the default approach when the \fct{tve} function is used to specify a time-varying effect for a covariate in the model formula. However, another approach is to use a piecewise constant function for modelling the time-varying log hazard ratio. If we want a piecewise constant log hazard ratio then we can specify \code{degree = 0} as an argument to the \fct{tve} function. This exploits the fact that the \fct{bSpline} function in the \pkg{splines2} package accepts \code{degree = 0} as a special case that corresponds to a piecewise constant basis.

We will again simulate some survival data using the \pkg{simsurv} package to show how a piecewise constant hazard ratio can be estimated using \fct{stan_surv}.

We simulate a dataset with $N = 1000$ individuals with event times generated under a Weibull hazard function with scale parameter $\lambda = 0.15$, shape parameter $\gamma = 1.1$, and binary baseline covariate $X_i \sim \text{Bern}(0.5)$. However, in this example our time-varying hazard ratio will be defined as $\beta(t) = -0.4 + 0.8 \times I(t > 4)$ where $I(x)$ is the indicator function taking the value 1 if $x$ is true and 0 otherwise. This corresponds to a piecewise constant log hazard ratio with just two "pieces" or time intervals. The first time interval is $[0,4]$ years during which the true hazard ratio is $\exp(-0.4) \approx 0.7$. The second time interval is $(4,\infty]$ years during which the true log hazard ratio is $\exp(-0.4 + 0.8) \approx 1.5$. Our example uses two time intervals for simplicity, but in general we could easily have considered more (although it would have required some additional lines of code to simulate the data). We will enforce administrative censoring at 15 years for those individuals whose simulated event time is >15 years:
\begin{Schunk}
\begin{Sinput}
R> set.seed(888222)
R> N <- 1000
R> covs <- data.frame(id  = 1:N, 
+                     trt = rbinom(N, 1, 0.5))
R> dat  <- simsurv(dist    = "weibull",
+                  lambdas = 0.15, 
+                  gammas  = 1.1, 
+                  x       = covs, 
+                  betas   = c(trt = -0.4),
+                  tde     = c(trt = 0.8),
+                  tdefun  = function(t) (t > 4),
+                  maxt    = 15)
R> dat  <- merge(dat, covs)
R> head(dat)
\end{Sinput}
\begin{Soutput}
  id eventtime status trt
1  1 3.7408753      1   0
2  2 4.7278398      1   1
3  3 1.3081087      1   1
4  4 0.1998067      1   1
5  5 1.1721203      1   0
6  6 0.2116944      1   0
\end{Soutput}
\end{Schunk}

We can estimate a model with a piecewise constant log hazard ratio for the covariate \code{trt} as follows:
\begin{Schunk}
\begin{Sinput}
R> mod3 <- stan_surv(
+    formula = Surv(eventtime, status) ~ tve(trt, degree = 0, knots = 4),
+    data    = dat,
+    basehaz = "weibull",
+    chains  = CHAINS, 
+    cores   = CORES, 
+    seed    = SEED,
+    iter    = ITER)
\end{Sinput}
\end{Schunk}

This time we specify some additional arguments to the \fct{tve} function so that our time-varying effect corresponds to the true data generating model used to simulate our event times. Specifically, we specify \code{degree = 0} to say that we want the time-varying effect (i.e. the time-varying log hazard ratio) to be estimated using a piecewise constant function and \code{knots = 4} to say that we only want one internal knot placed at time $t = 4$.

\begin{figure}[t!]
\centering
\includegraphics{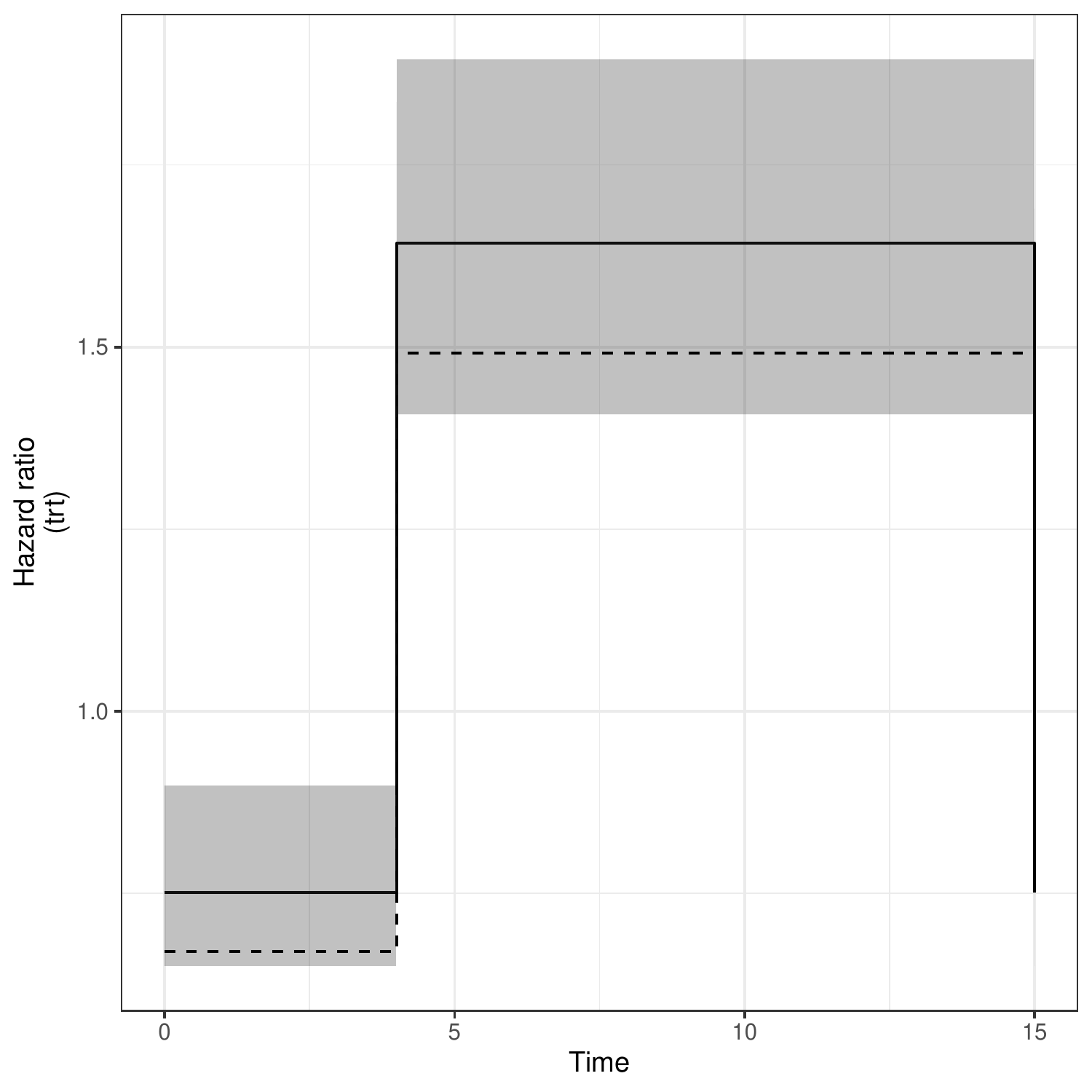}
\caption{\label{fig:tve-hr-pw} Time-varying hazard ratio for the estimated treatment effect (posterior median and 95\% uncertainty limits) when modelled using a piecewise constant function. The dashed line shows the "true" time-varying hazard ratio used to simulate the data.}
\end{figure}

We can again use the generic \fct{plot} method with argument \code{plotfun = "tve"} to visualise our estimated time-varying hazard ratio for treatment. This is shown in Figure \ref{fig:tve-hr-pw}. We see that the estimated hazard ratio reasonably reflects our true data generating model (i.e. a hazard ratio of $\approx 0.7$ during the first time interval and a hazard ratio of $\approx 1.5$ during the second time interval) although there is a slight discrepancy due to the sampling variation in the simulated event times.

\subsection{A multilevel survival model} \label{sec:multilevelmodel}

To demonstrate the estimation of a hierarchical model for survival data in \fct{stan_surv} we will use the \code{frail} dataset (see \code{help("rstanarm-datasets")} for a description). The \code{frail} dataset contains simulated event times for 200 patients clustered within 20 hospital sites (10 patients per hospital site). The event times are simulated from a parametric proportional hazards model under the following assumptions:

\begin{itemize}

\item a constant (i.e. exponential) baseline hazard rate of 0.1;

\item a fixed treatment effect with log hazard ratio of 0.3; and

\item a site-specific random intercept (specified on the log hazard scale) drawn from a $N(0,1)$ distribution.

\end{itemize}

Let's look at the first few rows of the data:
\begin{Schunk}
\begin{Sinput}
R> head(frail)
\end{Sinput}
\begin{Soutput}
  id site trt         b eventtime status
1  1    1   0 0.4229517 0.9058188      1
2  2    1   1 0.4229517 5.9190576      1
3  3    1   0 0.4229517 7.8525219      1
4  4    1   0 0.4229517 1.2066141      1
5  5    1   1 0.4229517 1.1703645      1
6  6    1   0 0.4229517 2.6209007      1
\end{Soutput}
\end{Schunk}

The dataset contains the unique patient identifier (\code{id}), unique site identifier (\code{site}), a treatment indicator (\code{trt}), the true value for the site-specific random effect (\code{b}), the event or censoring time (\code{eventtime}), and an event indicator (\code{status}).

To fit a hierarchical model for clustered survival data we use a formula syntax similar to what is used in the \pkg{lme4} \proglang{R} package \citep{Bates:2015}. Let's consider the following model (which aligns with the model used to generate the simulated data):
\begin{Schunk}
\begin{Sinput}
R> mod_randint <- stan_surv(
+    formula = Surv(eventtime, status) ~ trt + (1 | site),
+    data    = frail,
+    basehaz = "exp",
+    chains  = CHAINS, 
+    cores   = CORES, 
+    seed    = SEED,
+    iter    = ITER)
\end{Sinput}
\end{Schunk}

The model contains a baseline covariate for treatment (0 or 1) as well as a site-specific intercept to allow for correlation in the event times for patients from the same hospital site. We've called the model object \code{mod_randint} to denote the fact that it includes a site-specific (random) intercept. Let's examine the parameter estimates from the model:
\begin{Schunk}
\begin{Sinput}
R> print(mod_randint, digits = 2)
\end{Sinput}
\begin{Soutput}
stan_surv
 baseline hazard: exponential
 formula:         Surv(eventtime, status) ~ trt + (1 | site)
 observations:    200
 events:          152 (76%)
 right censored:  48 (24%)
 delayed entry:   no
------
            Median MAD_SD exp(Median)
(Intercept) -2.23   0.29     NA      
trt          0.45   0.17   1.57      

Error terms:
 Groups Name        Std.Dev.
 site   (Intercept) 1.12    
Num. levels: site 20 

------
* For help interpreting the printed output see ?print.stanreg
* For info on the priors used see ?prior_summary.stanreg
\end{Soutput}
\end{Schunk}

We see that the estimated log hazard ratio for treatment ($\hat{\beta}_{\text{(trt)}} = 0.45$) is about 50\% larger than the "true" log hazard ratio used in the data generating model ($\beta_{\text{(trt)}} = 0.3$). However the true value lies within +/- 1 posterior standard deviation of the posterior median so it is not incompatible with the data. The estimated baseline hazard rate is $\exp(-2.32) = 0.1$, which (to 1 d.p.) is equal to the baseline hazard rate used in the data generating model (0.1). Of course, slight differences between the estimated parameters and the true parameters from the data generating model are to be expected due to sampling variation.

If this were a real analysis, we might wonder whether the site-specific estimates are necessary. We can assess that by fitting an alternative model that does not include the site-specific intercepts and compare it to the model that does include them. We will compare it using the \fct{loo} function. We first need to fit the model without the site-specific intercept. To do this, we will just use the generic `update` method for `stansurv` objects, since all we are changing is the model formula:
\begin{Schunk}
\begin{Sinput}
R> mod_fixed <- update(
+    mod_randint, formula. = Surv(eventtime, status) ~ trt) 
\end{Sinput}
\end{Schunk}

Let's calculate the \code{loo} for both these models and compare them:
\begin{Schunk}
\begin{Sinput}
R> loo_fixed   <- loo(mod_fixed)
R> loo_randint <- loo(mod_randint)
R> compare_models(loo_fixed, loo_randint)
\end{Sinput}
\begin{Soutput}
Model formulas: 
 :  NULL
 :  NULLelpd_diff        se 
     56.7       9.6 
\end{Soutput}
\end{Schunk}

We see strong evidence in favour of the model with the site-specific intercepts.

What about if we want to generalise the random effects structure further? For instance, suppose we wish to know whether the site-specific intercept provides sufficient complexity. We can therefore consider estimating a model with both a site-specific intercept and a site-specific treatment effect.

The following code fits a model with a site-specific intercept and a site-specific coefficient for the covariate \code{trt} (i.e. treatment):
\begin{Schunk}
\begin{Sinput}
R> mod_randtrt <- update(
+    mod_randint, formula. = Surv(eventtime, status) ~ trt + (trt | site)) 
\end{Sinput}
\end{Schunk}
\begin{Schunk}
\begin{Sinput}
R> print(mod_randtrt, digits = 2)
\end{Sinput}
\begin{Soutput}
stan_surv
 baseline hazard: exponential
 formula:         Surv(eventtime, status) ~ trt + (trt | site)
 observations:    200
 events:          152 (76%)
 right censored:  48 (24%)
 delayed entry:   no
------
            Median MAD_SD exp(Median)
(Intercept) -2.24   0.28     NA      
trt          0.47   0.20   1.61      

Error terms:
 Groups Name        Std.Dev. Corr 
 site   (Intercept) 1.145         
        trt         0.447    -0.22
Num. levels: site 20 

------
* For help interpreting the printed output see ?print.stanreg
* For info on the priors used see ?prior_summary.stanreg
\end{Soutput}
\end{Schunk}

We see that we have an estimated standard deviation for the site-specific intercepts and the site-specific coefficients for \code{trt}, as well as the estimated correlation between those site-specific parameters.

Let's now compare all three of these models based on \code{loo}:
\begin{Schunk}
\begin{Sinput}
R> loo_randtrt <- loo(mod_randtrt)
R> compare_models(loo_fixed, loo_randint, loo_randtrt)
\end{Sinput}
\begin{Soutput}
Model formulas: 
 :  NULL
 :  NULL
 :  NULL            elpd_diff se_diff
mod_randint   0.0       0.0  
mod_randtrt  -1.0       0.8  
mod_fixed   -56.7       9.6  
\end{Soutput}
\end{Schunk}

It appears that the model with just a site-specific intercept is the best fitting model. It is much better than the model without a site-specific intercept, and slightly better than the model with both a site-specific intercept and a site-specific treatment effect. In other words, including a site-specific intercept appears important, but including a site-specific treatment effect is not. This conclusion is reassuring because it aligns with the data generating model we used to simulate the data.


\section{Summary} \label{sec:summary}

The \pkg{rstanarm} \proglang{R} package provides a suite of functions for applied Bayesian regression modelling and it has recently been extended to include survival models. The syntax for fitting survival models in \pkg{rstanarm} is user-friendly and built around customary \proglang{R} formulas and data frames. There is a broad range of survival models currently accommodated. Time-varying covariates, time-varying effects, multilevel survival models, all manners of censoring (left, right, interval) and delayed entry (left truncation) can all be handled. These features allow us to overcome many of the challenges encountered in analysing "real world" survival data. A range of choices for prior distributions are provided and they allow significant flexibility from a Bayesian modelling perspective.

A number of extensions are still possible such as competing risks, recurrent events, and multi-state models. Alternative baseline hazards based on penalised splines or non-parametric Gaussian processes could also be used. We plan to make several of these additions in the future.

One significant extension that is already available is the joint modelling of longitudinal (e.g. a repeatedly measured clinical biomarker) and survival data. Joint models for longitudinal and survival data are implemented via the \fct{stan_jm} function in \pkg{rstanarm} \citep{Brilleman:2018}. The methodology underpinning joint modelling is a significant extension beyond the relatively standard survival models described in this article. For that reason we believe the \fct{stan_jm} modelling function is outside the scope of this paper and we have chosen to document it elsewhere. Nonetheless joint models for longitudinal and survival data are a branch of survival models and individuals who wish to consider the association between a time-varying covariate and a survival endpoint could consider whether the \fct{stan_jm} modelling function in \pkg{rstanarm} may be appropriate for their context.

We hope that the inclusion of survival modelling functionality in \pkg{rstanarm} will help to increase the uptake of Bayesian survival analysis in applied research.


\section*{Acknowledgments}

SLB is funded by an Australian National Health and Medical Research Council (NHMRC) Project Grant (ref: 1128222).


\bibliography{refs}


\newpage

\begin{appendix}

\section{Parameterisations on the hazard scale} \label{app:haz-parameterisations}

When \code{basehaz} is set equal to \code{"exp"}, \code{"weibull"}, \code{"gompertz"}, \code{"ms"} (the default), or \code{"bs"} then the model is defined on the hazard scale using the following parameterisations. We first introduce each parameterisation under the assumption of a time-fixed linear predictor and then in Section \ref{sec:param-haz-tve} we show the extension to time-varying effects.

\subsection{Exponential model}

The exponential model is parameterised with scale parameter $\lambda_i = \exp(\eta_i)$.

For individual $i$ we have:
\begin{align}
\begin{split}
  h_i(T_i) 
    & = \lambda_i \\
    & = \exp(\eta_i) \\
  H_i(T_i) 
    & = T_i \lambda_i \\
    & = T_i \exp(\eta_i) \\
  S_i(T_i) 
    & = \exp \left( - T_i \lambda_i \right) \\
    & = \exp \left( - T_i \exp(\eta_i) \right) \\
  F_i(T_i) 
    & = 1 - \exp \left( - T_i \lambda_i \right) \\
    & = 1 - \exp \left( - T_i \exp(\eta_i) \right) \\
  S_i(T_i) - S_i(T_i^U) 
    & = \exp \left( - T_i \lambda_i \right) - \exp \left( - T_i^U \lambda_i \right) \\
    & = \exp \left( - T_i \exp(\eta_i) \right) - \exp \left( - T_i^U \exp(\eta_i) \right)
\end{split}
\end{align}
or on the log scale:
\begin{align}
\begin{split}
  \log h_i(T_i) 
    & = \log \lambda_i \\
    & = \eta_i \\
  \log H_i(T_i) 
    & = \log(T_i) + \log \lambda_i \\
    & = \log(T_i) + \eta_i \\
  \log S_i(T_i) 
    & = - T_i \lambda_i \\
    & = - T_i \exp(\eta_i) \\
  \log F_i(T_i) 
    & = \log \left( 1 - \exp \left( - T_i \lambda_i \right) \right) \\
    & = \log \left( 1 - \exp \left( - T_i \exp(\eta_i) \right) \right) \\
  \log (S_i(T_i) - S_i(T_i^U)) 
    & = \log \left[ \exp \left( - T_i \lambda_i \right) - \exp \left( - T_i^U \lambda_i \right) \right] \\
    & = \log \left[ \exp \left( - T_i \exp(\eta_i) \right) - \exp \left( - T_i^U \exp(\eta_i) \right) \right]
\end{split}
\end{align}
  
\subsection{Weibull model}

The Weibull model is parameterised with scale parameter $\lambda_i = \exp(\eta_i)$ and shape parameter $\gamma > 0$. 

For individual $i$ we have:
\begin{align}
\begin{split}
  h_i(T_i) 
    & = \gamma T_i^{\gamma-1} \lambda_i \\
    & = \gamma T_i^{\gamma-1} \exp(\eta_i) \\
  H_i(T_i) 
    & = T_i^{\gamma} \lambda_i \\
    & = T_i^{\gamma} \exp(\eta_i) \\
  S_i(T_i) 
    & = \exp \left( - T_i^{\gamma} \lambda_i \right) \\
    & = \exp \left( - T_i^{\gamma} \exp(\eta_i) \right) \\
  F_i(T_i) 
    & = 1 - \exp \left( - T_i^{\gamma} \lambda_i \right) \\
    & = 1 - \exp \left( - T_i^{\gamma} \exp(\eta_i) \right) \\
  S_i(T_i) - S_i(T_i^U) 
    & = \exp \left( - T_i^{\gamma} \lambda_i \right) - \exp \left( - T_i^{U \gamma} \lambda_i \right) \\
    & = \exp \left( - T_i^{\gamma} \exp(\eta_i) \right) - \exp \left( - T_i^{U \gamma} \exp(\eta_i) \right)
\end{split}
\end{align}
or on the log scale:
\begin{align}
\begin{split}
  \log h_i(T_i) 
    & = \log(\gamma) + (\gamma-1) \log(t) + \log \lambda_i \\
    & = \log(\gamma) + (\gamma-1) \log(t) + \eta_i \\
  \log H_i(T_i) 
    & = \gamma \log(T_i) + \log \lambda_i \\
    & = \gamma \log(T_i) + \eta_i \\
  \log S_i(T_i) 
    & = - T_i^{\gamma} \lambda_i \\
    & = - T_i^{\gamma} \exp(\eta_i) \\
  \log F_i(T_i) 
    & = \log \left( 1 - \exp \left( - T_i^{\gamma} \lambda_i \right) \right) \\
    & = \log \left( 1 - \exp \left( - T_i^{\gamma} \exp(\eta_i) \right) \right) \\
  \log (S_i(T_i) - S_i(T_i^U)) 
    & = \log \left[ \exp \left( - T_i^{\gamma} \lambda_i \right) - \exp \left( - T_i^{U \gamma} \lambda_i \right) \right] \\
    & = \log \left[ \exp \left( - T_i^{\gamma} \exp(\eta_i) \right) - \exp \left( - T_i^{U \gamma} \exp(\eta_i) \right) \right]
\end{split}
\end{align}

\subsection{Gompertz model}

The Gompertz model is parameterised with shape parameter $\lambda_i = \exp(\eta_i)$ and scale parameter $\gamma > 0$. 

For individual $i$ we have:
\begin{align}
\begin{split}
  h_i(T_i) 
    & = \exp(\gamma T_i) \lambda_i \\
    & = \exp(\gamma T_i) \exp(\eta_i) \\
  H_i(T_i) 
    & = \frac{\exp(\gamma T_i) - 1}{\gamma} \lambda_i \\
    & = \frac{\exp(\gamma T_i) - 1}{\gamma} \exp(\eta_i) \\
  S_i(T_i) 
    & = \exp \left( \frac{-(\exp(\gamma T_i) - 1)}{\gamma} \lambda_i \right) \\
    & = \exp \left( \frac{-(\exp(\gamma T_i) - 1)}{\gamma} \exp(\eta_i) \right) \\
  F_i(T_i) 
    & = 1 - \exp \left( \frac{-(\exp(\gamma T_i) - 1)}{\gamma} \lambda_i \right) \\
    & = 1 - \exp \left( \frac{-(\exp(\gamma T_i) - 1)}{\gamma} \exp(\eta_i) \right) \\
  S_i(T_i) - S_i(T_i^U) 
    & = \exp \left( \frac{-(\exp(\gamma T_i) - 1)}{\gamma} \lambda_i \right) - \exp \left( \frac{-(\exp(\gamma T_i^U) - 1)}{\gamma} \lambda_i \right) \\
    & = \exp \left( \frac{-(\exp(\gamma T_i) - 1)}{\gamma} \exp(\eta_i) \right) - \exp \left( \frac{-(\exp(\gamma T_i^U) - 1)}{\gamma} \exp(\eta_i) \right)
\end{split}
\end{align}
or on the log scale:
\begin{align}
\begin{split}
  \log h_i(T_i) 
    & = \gamma T_i + \log \lambda_i \\
    & = \gamma T_i + \eta_i \\
  \log H_i(T_i) 
    & = \log(\exp(\gamma T_i) - 1) - \log(\gamma) + \log \lambda_i \\
    & = \log(\exp(\gamma T_i) - 1) - \log(\gamma) + \eta_i \\
  \log S_i(T_i) 
    & = \frac{-(\exp(\gamma T_i) - 1)}{\gamma} \lambda_i \\
    & = \frac{-(\exp(\gamma T_i) - 1)}{\gamma} \exp(\eta_i) \\
  \log F_i(T_i) 
    & = \log \left( 1 - \exp \left( \frac{-(\exp(\gamma T_i) - 1)}{\gamma} \lambda_i \right) \right) \\
    & = \log \left( 1 - \exp \left( \frac{-(\exp(\gamma T_i) - 1)}{\gamma} \exp(\eta_i) \right) \right) \\
  \log (S_i(T_i) - S_i(T_i^U)) 
    & = \log \left[ \exp \left( \frac{-(\exp(\gamma T_i) - 1)}{\gamma} \lambda_i \right) - \exp \left( \frac{-(\exp(\gamma T_i^U) - 1)}{\gamma} \lambda_i \right) \right] \\
    & = \log \left[ \exp \left( \frac{-(\exp(\gamma T_i) - 1)}{\gamma} \exp(\eta_i) \right) - \exp \left( \frac{-(\exp(\gamma T_i^U) - 1)}{\gamma} \exp(\eta_i) \right) \right]
\end{split}
\end{align}

\subsection{M-spline model}

Following on from Section \ref{sec:modelformulations} in the main text, let the M-spline function be denoted $M(t; \boldsymbol{\gamma}, \boldsymbol{k}, \delta) = \sum_{l=1}^{L} \gamma_{l} M_{l}(t; \boldsymbol{k}, \delta)$ where $\boldsymbol{\gamma} > 0$ is the vector of M-spline coefficients for the baseline hazard.

Similarly, let $I(t; \boldsymbol{\gamma}, \boldsymbol{k}, \delta) = \sum_{l=1}^{L} \gamma_{l} I_{l}(t; \boldsymbol{k}, \delta)$ denote the corresponding I-spline function (i.e. integral of an M-spline) evaluated using the same degree $\delta$, knot locations $\boldsymbol{k}$, and coefficients $\boldsymbol{\gamma}$.

Note that both the M-spline and I-spline functions can be evaluated analytically with the basis terms $M_{l}(t; \boldsymbol{k}, \delta)$ or $I_{l}(t; \boldsymbol{k}, \delta)$ for $l = 1,...,L$ calculated using the \fct{mSpline} and \fct{iSpline} functions in the \pkg{splines2} \proglang{R} package \citep{Wang:2018}.

For individual $i$ we have:
\begin{align}
\begin{split}
  h_i(T_i) 
    & = M(T_i; \boldsymbol{\gamma}, \boldsymbol{k}, \delta) \exp(\eta_i) \\
  H_i(T_i) 
    & = I(T_i; \boldsymbol{\gamma}, \boldsymbol{k}, \delta) \exp(\eta_i) \\
  S_i(T_i) 
    & = \exp \left( - I(T_i; \boldsymbol{\gamma}, \boldsymbol{k}, \delta) \exp(\eta_i) \right) \\
  F_i(T_i) 
    & = 1 - \exp \left( - I(T_i; \boldsymbol{\gamma}, \boldsymbol{k}, \delta) \exp(\eta_i) \right) \\
  S_i(T_i) - S_i(T_i^U) 
    & = \exp \left( - I(T_i; \boldsymbol{\gamma}, \boldsymbol{k}, \delta) \exp(\eta_i) \right) - \exp \left( - I(T_i^U; \boldsymbol{\gamma}, \boldsymbol{k}, \delta) \exp(\eta_i) \right)
\end{split}
\end{align}
or on the log scale:
\begin{align}
\begin{split}
  \log h_i(T_i) 
    & = \log(M(T_i; \boldsymbol{\gamma}, \boldsymbol{k}, \delta)) + \eta_i \\
  \log H_i(T_i) 
    & = \log(I(T_i; \boldsymbol{\gamma}, \boldsymbol{k}, \delta)) + \eta_i \\
  \log S_i(T_i) 
    & = - I(T_i; \boldsymbol{\gamma}, \boldsymbol{k}, \delta) \exp(\eta_i) \\
  \log F_i(T_i) 
    & = \log \left[ 1 - \exp \left( - I(T_i; \boldsymbol{\gamma}, \boldsymbol{k}, \delta) \exp(\eta_i) \right) \right] \\
  \log (S_i(T_i) - S_i(T_i^U)) 
    & = \log \left[ \exp \left( - I(T_i; \boldsymbol{\gamma}, \boldsymbol{k}, \delta) \exp(\eta_i) \right) - \exp \left( - I(T_i^U; \boldsymbol{\gamma}, \boldsymbol{k}, \delta) \exp(\eta_i) \right) \right]
\end{split}
\end{align}

\subsection{B-spline model}

Following on from Section \ref{sec:modelformulations} in the main text, let the B-spline function be denoted $B(t; \boldsymbol{\gamma}, \boldsymbol{k}, \delta) = \sum_{l=1}^{L} \gamma_{l} B_{l}(t; \boldsymbol{k}, \delta)$ where $\boldsymbol{\gamma}$ is the vector of B-spline coefficients for the log baseline hazard.

Note that both the B-spline function can be evaluated analytically with the basis terms $B_{l}(t; \boldsymbol{k}, \delta)$ for $l = 1,...,L$ calculated using the \fct{bSpline} in the \pkg{splines2} \proglang{R} package \citep{Wang:2018}.

For individual $i$ we have:
\begin{align}
\begin{split}
  h_i(T_i) 
    & = \exp \left( B(T_i; \boldsymbol{\gamma}, \boldsymbol{k}, \delta) + \eta_i \right)
\end{split}
\end{align}
or on the log scale:
\begin{align}
\begin{split}
  \log h_i(T_i) 
    & = B(T_i; \boldsymbol{\gamma}, \boldsymbol{k}, \delta) + \eta_i
\end{split}
\end{align}

The cumulative hazard, survival function, and CDF for the B-spline model cannot be calculated analytically. Instead, the model is only defined analytically on the hazard scale and Gauss-Kronrod quadrature (see Section \ref{sec:loglikelihood} of the main text) is used to evaluate the following:
\begin{align}
\begin{split}
  H_i(T_i) 
    & = \int_{u=0}^{T_i} h_i(u) du \\
  S_i(T_i) 
    & = \exp \left( - \int_{u=0}^{T_i} h_i(u) du \right) \\
  F_i(T_i) 
    & = 1 - \exp \left( - \int_{u=0}^{T_i} h_i(u) du \right) \\
  S_i(T_i) - S_i(T_i^U) 
    & = \exp \left( -\int_{u=0}^{T_i} h_i(u) du \right) - \exp \left( - \int_{u=0}^{T_i^U} h_i(u) du \right)
\end{split}
\end{align}

\subsection{Extension to time-varying effects (i.e. non-proportional hazards)} \label{sec:param-haz-tve}

We can extend the previous model formulations to allow for time-varying coefficients (i.e. non-proportional hazards). The time-varying linear predictor is introduced on the hazard scale. That is, $\eta_i$ in our previous model definitions is instead replaced by $\eta_i(t)$. This leads to an analytical form for the hazard and log hazard. However, in general, there is no longer a closed form expression for the cumulative hazard, survival function, or CDF. Therefore, when the linear predictor includes time-varying coefficients, quadrature is used to evaluate the following:
\begin{align}
\begin{split}
  H_i(T_i) 
    & = \int_{u=0}^{T_i} h_i(u) du \\
  S_i(T_i) 
    & = \exp \left( - \int_{u=0}^{T_i} h_i(u) du \right) \\
  F_i(T_i) 
    & = 1 - \exp \left( - \int_{u=0}^{T_i} h_i(u) du \right) \\
  S_i(T_i) - S_i(T_i^U) 
    & = \exp \left( -\int_{u=0}^{T_i} h_i(u) du \right) - \exp \left( - \int_{u=0}^{T_i^U} h_i(u) du \right)
\end{split}
\end{align}

\section{Parameterisations under accelerated failure times} \label{app:aft-parameterisations}

When \code{basehaz} is set equal to \code{"exp-aft"} or \code{"weibull-aft"} then the model is defined on the accelerated failure time (AFT) scale using the following parameterisations. We first introduce each parameterisation under the assumption of a time-fixed linear predictor and then in Section \ref{sec:param-aft-tve} we show the extension to time-varying effects.

\subsection{Exponential model}

The exponential model is parameterised with scale parameter $\lambda_i = \exp(-\eta_i)$.

For individual $i$ we have:
\begin{align}
\begin{split}
  h_i(T_i) 
    & = \lambda_i \\
    & = \exp(-\eta_i) \\
  H_i(T_i) 
    & = T_i \lambda_i \\ 
    & = T_i \exp(-\eta_i) \\
  S_i(T_i) 
    & = \exp \left( - T_i \lambda_i \right) \\
    & = \exp \left( - T_i \exp(-\eta_i) \right) \\
  F_i(T_i) 
    & = 1 - \exp \left( - T_i \lambda_i \right) \\
    & = 1 - \exp \left( - T_i \exp(-\eta_i) \right) \\
  S_i(T_i) - S_i(T_i^U) 
    & = \exp \left( - T_i \lambda_i \right) - \exp \left( - T_i^U \lambda_i \right) \\
    & = \exp \left( - T_i \exp(-\eta_i) \right) - \exp \left( - T_i^U \exp(-\eta_i) \right)
\end{split}
\end{align}
or on the log scale:
\begin{align}
\begin{split}
  \log h_i(T_i) 
    & = \log \lambda_i \\
    & = -\eta_i \\
  \log H_i(T_i) 
    & = \log(T_i) + \log \lambda_i \\
    & = \log(T_i) - \eta_i \\
  \log S_i(T_i) 
    & = - T_i \lambda_i \\
    & = - T_i \exp(-\eta_i) \\
  \log F_i(T_i) 
    & = \log \left( 1 - \exp \left( - T_i \lambda_i \right) \right) \\
    & = \log \left( 1 - \exp \left( - T_i \exp(-\eta_i) \right) \right) \\
  \log (S_i(T_i) - S_i(T_i^U)) 
    & = \log \left[ \exp \left( - T_i \lambda_i) \right) - \exp \left( - T_i^U \lambda_i \right) \right] \\
    & = \log \left[ \exp \left( - T_i \exp(-\eta_i) \right) - \exp \left( - T_i^U \exp(-\eta_i) \right) \right]
\end{split}
\end{align}

Note that Section \ref{sec:linearpredictor} of the main text described the relationship between regression coefficients from an exponential proportional hazards model and an exponential AFT model.

Lastly, note that the general form for the hazard and survival functions under an AFT model with acceleration factor $\exp(-\eta_i)$ can be used to derive the exponential AFT model defined here by setting $h_0(t) = 1$, $S_0(t) = \exp(-t)$, and $\lambda_i = \exp(-\eta_i)$:
\begin{align}
\begin{split}
  h_i(T_i) 
    & = \exp(-\eta_i) h_0(T_i \exp(-\eta_i)) \\
    & = \exp(-\eta_i) \\
    & = \lambda_i
\end{split}
\end{align}
\begin{align}
\begin{split}
  S_i(T_i) 
    & = S_0(T_i \exp(-\eta_i)) \\
    & = \exp(-T_i \exp(-\eta_i)) \\
    & = \exp(-T_i \lambda_i)
\end{split}
\end{align}

\subsection{Weibull model}

The Weibull model is parameterised with scale parameter $\lambda_i = \exp(-\gamma \eta_i)$ and shape parameter $\gamma > 0$.

For individual $i$ we have:
\begin{align}
\begin{split}
  h_i(T_i) 
    & = \gamma T_i^{\gamma-1} \lambda_i \\
    & = \gamma T_i^{\gamma-1} \exp(-\gamma \eta_i) \\
  H_i(T_i) 
    & = T_i^{\gamma} \lambda_i \\
    & = T_i^{\gamma} \exp(-\gamma \eta_i) \\
  S_i(T_i) 
    & = \exp \left( - T_i^{\gamma} \lambda_i \right) \\
    & = \exp \left( - T_i^{\gamma} \exp(-\gamma \eta_i) \right) \\
  F_i(T_i) 
    & = 1 - \exp \left( - T_i^{\gamma} \lambda_i \right) \\
    & = 1 - \exp \left( - T_i^{\gamma} \exp(-\gamma \eta_i) \right) \\
  S_i(T_i) - S_i(T_i^U) 
    & = \exp \left( - T_i^{\gamma} \lambda_i \right) - \exp \left( - T_i^{U \gamma} \lambda_i \right) \\
    & = \exp \left( - T_i^{\gamma} \exp(-\gamma \eta_i) \right) - \exp \left( - T_i^{U \gamma} \exp(-\gamma \eta_i) \right)
\end{split}
\end{align}
or on the log scale:
\begin{align}
\begin{split}
  \log h_i(T_i) 
    & = \log(\gamma) + (\gamma-1) \log(T_i) + \log \lambda_i \\
    & = \log(\gamma) + (\gamma-1) \log(T_i) - \gamma \eta_i \\
  \log H_i(T_i) 
    & = \gamma \log(T_i) + \log \lambda_i \\
    & = \gamma \log(T_i) - \gamma \eta_i \\
  \log S_i(T_i) 
    & = - T_i^{\gamma} \lambda_i \\
    & = - T_i^{\gamma} \exp(-\gamma \eta_i) \\
  \log F_i(T_i) 
    & = \log \left( 1 - \exp \left( - T_i^{\gamma} \lambda_i \right) \right) \\
    & = \log \left( 1 - \exp \left( - T_i^{\gamma} \exp(-\gamma \eta_i) \right) \right) \\
  \log (S_i(T_i) - S_i(T_i^U)) 
    & = \log \left[ \exp \left( - T_i^{\gamma} \lambda_i \right) - \exp \left( - T_i^{U \gamma} \lambda_i \right) \right] \\
    & = \log \left[ \exp \left( - T_i^{\gamma} \exp(-\gamma \eta_i) \right) - \exp \left( - T_i^{U \gamma} \exp(-\gamma \eta_i) \right) \right]
\end{split}
\end{align}

Note that Section \ref{sec:linearpredictor} of the main text described the relationship between regression coefficients from a Weibull proportional hazards model and a Weibull AFT model.

Lastly, note that the general form for the hazard and survival functions under an AFT model with acceleration factor $\exp(-\eta_i)$ can be used to derive the Weibull AFT model defined here by setting $h_0(t) = \gamma t^{\gamma - 1}$, $S_0(t) = \exp(-t^{\gamma})$, and $\lambda_i = \exp(-\gamma \eta_i)$:
\begin{align}
\begin{split}
  h_i(T_i) 
    & = \exp(-\eta_i) h_0(T_i \exp(-\eta_i)) \\
    & = \exp(-\eta_i) \gamma {(T_i \exp(-\eta_i))}^{\gamma - 1} \\
    & = \exp(-\gamma \eta_i) \gamma T_i^{\gamma - 1} \\
    & = \lambda_i \gamma T_i^{\gamma - 1}
\end{split}
\end{align}
\begin{align}
\begin{split}
  S_i(T_i) 
    & = S_0(T_i \exp(-\eta_i)) \\
    & = \exp(-(T_i \exp(-\eta_i))^{\gamma}) \\
    & = \exp(-T_i^{\gamma} [\exp(-\eta_i)]^{\gamma}) \\
    & = \exp(-T_i^{\gamma} \exp(-\gamma \eta_i)) \\
    & = \exp(-T_i \lambda_i)
\end{split}
\end{align}

\subsection{Extension to time-varying coefficients (i.e. time-varying acceleration factors)} \label{sec:param-aft-tve}

We can extend the previous model formulations to allow for time-varying coefficients (i.e. time-varying acceleration factors). 

The so-called "unmoderated" survival probability for an individual at time $t$ is defined as the baseline survival probability at time $t$, i.e. $S_i(t) = S_0(t)$. With a time-fixed acceleration factor, the survival probability for a so-called "moderated" individual is defined as the baseline survival probability but evaluated at "time $t$ multiplied by the acceleration factor $\exp(-\eta_i)$". That is, the survival probability for the moderated individual is $S_i(t) = S_0(t \exp(-\eta_i))$. 

However, with time-varying acceleration we cannot simply multiply time by a fixed (acceleration) constant. Instead, we must integrate the function for the time-varying acceleration factor over the interval $0$ to $t$. In other words, we must evaluate:
\begin{align}
\begin{split}
  S_i(t) = S_0 \left( \int_{u=0}^t \exp(-\eta_i(u)) du \right)
\end{split}
\end{align}
as described by Hougaard \cite{Hougaard:1999}.

Hougaard also gives a general expression for the hazard function under time-varying acceleration, as follows:
\begin{align}
\begin{split}
  h_i(t) = \exp \left(-\eta_i(t) \right) h_0 \left( \int_{u=0}^t \exp(-\eta_i(u)) du \right)
\end{split}
\end{align}

It is interesting to note here that the hazard at time $t$ is in fact a function of the full history of covariates and parameters (i.e. the linear predictor) from time $0$ up until time $t$. This is different to the hazard scale formulation of time-varying effects (i.e. non-proportional hazards). Under the hazard scale formulation with time-varying effects, the survival probability is a function of the full history between times $0$ and $t$, but the hazard is not. Instead, under a hazard scale formulation the hazard rate is only a function of the current value of the covariate(s) and parameter(s). This is particularly important to consider when fitting AFT models with time-varying effects in the presence of delayed entry (i.e. left truncation).

For the exponential distribution, this leads to:
\begin{align}
\begin{split}
  S_i(T_i) 
    & = S_0 \left( \int_{u=0}^{T_i} \exp(-\eta_i(u)) du \right) \\
    & = \exp \left(- \int_{u=0}^{T_i} \exp(-\eta_i(u)) du \right)
\end{split}
\end{align}
\begin{align}
\begin{split}
  h_i(T_i) 
    & = \exp \left(-\eta_i(T_i) \right) h_0 \left( \int_{u=0}^{T_i} \exp(-\eta_i(u)) du \right) \\
    & = \exp \left(-\eta_i(T_i) \right) \exp \left(- \int_{u=0}^{T_i} \exp(-\eta_i(u)) du \right)
\end{split}
\end{align}
and for the Weibull distribution, this leads to:
\begin{align}
\begin{split}
  S_i(T_i) 
    & = S_0 \left( \int_{u=0}^{T_i} \exp(-\eta_i(u)) du \right) \\
    & = \exp \left(- \left[\int_{u=0}^{T_i} \exp (-\eta_i(u)) du \right]^{\gamma} \right)
\end{split}
\end{align}
\begin{align}
\begin{split}
  h_i(T_i) 
    & = \exp \left(-\eta_i(T_i) \right) h_0 \left( \int_{u=0}^{T_i} \exp(-\eta_i(u)) du \right) \\
    & = \exp \left(-\eta_i(T_i) \right) \exp \left(- \left[\int_{u=0}^{T_i} \exp (-\eta_i(u)) du \right]^{\gamma} \right)
\end{split}
\end{align}

These general expressions for the hazard and survival function under an AFT model with a time-varying linear predictor are used to evaluate the likelihood for the AFT model in \fct{stan_surv} when time-varying effects are specified in the model formula. Specifically, Gauss-Kronrod quadrature is used to evaluate the cumulative acceleration factor $\int_{u=0}^t \exp(-\eta_i(u)) du$ and this is then substituted into the relevant expressions for the hazard and survival.

\end{appendix}


\end{document}